\documentclass[manuscript]{acmart}

\AtBeginDocument{%
  }

\setcopyright{acmcopyright}
\copyrightyear{2024}
\acmYear{2024}
\acmDOI{10.1145/3613904.3642824}

\acmConference[CHI'24]{Make sure to enter the correct
  conference title from your rights confirmation emai}{May 11--16,
  2024}{Honolulu, HI, USA}

\acmSubmissionID{5741}

\usepackage{enumitem}
\usepackage{amsmath}
\usepackage{multirow}
\usepackage{xspace}
\usepackage[normalem]{ulem}
\usepackage{url}

\usepackage[nonumberlist,nopostdot]{glossaries}
\setglossarystyle{altlist}
\makenoidxglossaries
\newglossaryentry{generative artificial intelligence}
{
        name=Generative Artificial Intelligence (GAI),
        description={Artificial intelligence can generate new content (e.g., text and images) through training on an extensive dataset~\cite{generativeAI}.}
}

\newglossaryentry{Large Language Model}
{
        name=Large Language Model (LLM),
        text=large language model (LLM),
        description={The language model can comprehend and produce human-like responses based on a large amount of textual data training.}
}

\newglossaryentry{artificial intelligence generated content}
{
        name=Artificial Intelligence Generated Content (AIGC),
        text=AIGC,
        description={A new creation model incorporating artificial intelligence and human collaboration for content generation.}
}

\newglossaryentry{graphical user interface (GUI)}
{
        name=Graphical User Interface (GUI),
        text=graphical user interface (GUI),
        description={An interface that provides visual information and allows interaction.}
}

\newglossaryentry{prompts}
{
        name=Prompts,
        text=prompts,
        description={Instructions or requests given to generative artificial intelligence.}
}

\newglossaryentry{pre-trained model}
{
        name=Pre-trained text-to-image generation models,
        text=Pre-trained text-to-image generation models,
        description={Generative artificial intelligence (GAI) models that have been previously trained can be further tuned to perform text-to-image tasks.}
}

\newglossaryentry{finetune}
{
        name=Fine-tune,
        text=fine-tuned,
        description={A technique to train a pre-trained model to optimize its performance on a specific task.}
}

\newglossaryentry{lora}
{
        name=Low-Rank Adaptation (LoRA) Diffusion Model,
        text=low-rank adaptation diffusion model,
        description={A method of fine-tuning large pre-trained models for text-to-image generation enables the training with fewer parameters while ensuring high performance, reducing computational cost, and allowing large models to excel in specific tasks~\cite{hu2021lora}.}
}

\newglossaryentry{endtoend}
{
        name=End to End,
        text=end-to-end,
        description={An idea that only focuses on inputs and outputs without considering intermediate processes.}
}

\newglossaryentry{latent diffusion model}
{
        name=Latent Diffusion Model,
        text=latent diffusion model,
        description={Based on the traditional diffusion model, this model applies the diffusion process to the \gls{latent space} encoding rather than the pixels, thus generating high-quality images more efficiently\cite{rombach2022high,li2023gligen}.}
}

\newglossaryentry{latent space}
{
        name=Latent Space,
        text=latent,
        description={It is a conceptual space of highly compressed data with universal characteristics, which can be used to interpret the internal features and laws of the data effectively.}
}

\newglossaryentry{zeroshot}
{
        name=Zero shot,
        text=Zero-shot,
        description={A training approach using a model directly to perform tasks without additional training.}
}

\newglossaryentry{fewshot}
{
        name=Few shots,
        text=Few-shot,
        description={Another method to train models for specific tasks with a few samples.}
}

\newglossaryentry{boundingbox}
{
        name= Bounding box,
        text=bounding boxes,
        description={A rectangular box represents an object's spatial location and dimensional information.}
}

\newglossaryentry{promptengineering}
{
        name=Prompt Engineering,
        text=prompt engineering,
        description={One practice is to modify the prompts given to the large language models for better outputs that meet user needs.}
}

\newglossaryentry{UNet}
{
        name= Stable Diffusion UNet,
        text=Stable Diffusion UNet,
        description={A symmetrical structure of the central component in the stable diffusion model plays a crucial role in effectively denoising images.}
}

\newglossaryentry{gated}
{
        name= Gated Attention,
        text=gated-attention,
        description={It is a component of neural networks referred to as the attention mechanism that plays a crucial role in selectively focusing on targeted regions, thereby enhancing the model's ability to capture and extract key features.}
}

\newglossaryentry{DDIM}
{
        name=Denoising Diffusion Implicit Model (DDIM) inversion,
        text=DDIM inversion,
        description={A technique is proposed for iteratively transforming data distributions, such as images, into random noise. It enables the model to reconstruct or generate specific instances of data, such as images, by retracing back its corresponding latent space in the diffusion process.}
}

\newglossarystyle{twocolborderbold}{
  
    {
    \begin{longtable}{p{0.3\textwidth}p{0.65\textwidth}}
    \noalign{\hrule height 1.5pt} 
    }
    {
    \end{longtable}
    }%
  \renewcommand*{\subglossentry}[3]{%
  }%
}

\begin{document}

\newcommand{\eg}{\emph{e.g.}}
\newcommand{\ie}{\emph{i.e.}}

\newcommand{\strike}[1]{\textcolor{red}{\sout{#1}}}
\newcommand{\strikeg}[1]{\textcolor{blue}{\sout{#1}}}
\newcommand{\add}[1]{\textcolor{red}{#1}}
\newcommand{\replace}[2]{\strikeg{#1 }\add{#2}}
\newcommand{\zw}[1]{\textcolor{red}{Zeng: #1}}
\newcommand{\new}[1]{\textcolor{blue}{#1}}

\newcommand{\tool}{\emph{PlantoGraphy}\xspace}

\definecolor{newgreen}{rgb}{0.0, 0.5, 0.0}
\definecolor{newblue}{rgb}{0.0, 0.0, 0.0}
\definecolor{newred}{rgb}{1.0, 0.0, 0.0}

\definecolor{blue}{rgb}{0.0, 0.0, 0.0}
\newcommand{\rev}[1]{\textcolor{blue}{#1}}

\newenvironment{tightcenter}{%
  \setlength\topsep{0pt}
  \setlength\parskip{0pt}
  \begin{center}
}{%
  \end{center}
}

\title[PlantoGraphy]{PlantoGraphy: Incorporating Iterative Design Process into Generative Artificial Intelligence for Landscape Rendering
}


\author{Rong Huang}
\affiliation{%
  \institution{The Hong Kong University of Science and Technology (Guangzhou)}
  \city{Guangzhou}
  \country{China}
}

\author{Hai-Chuan Lin}
\affiliation{%
  \institution{The Hong Kong University of Science and Technology (Guangzhou)}
  \city{Guangzhou}
  \country{China}
}

\author{Chuanzhang Chen}
\affiliation{%
  \institution{The Hong Kong University of Science and Technology (Guangzhou)}
  \city{Guangzhou}
  \country{China}
}

\author{Kang Zhang}
\affiliation{%
  \institution{The Hong Kong University of Science and Technology (Guangzhou)}
  \city{Guangzhou}
  \country{China}
}
\affiliation{%
  \institution{The Hong Kong University of Science and Technology}
  \city{Hong Kong}
  \country{China}
}

\author{Wei Zeng}
\authornote{Wei Zeng is the corresponding author.}
\orcid{0000-0002-5600-8824}
\affiliation{%
  \institution{The Hong Kong University of Science and Technology (Guangzhou)}
  \city{Guangzhou}
  \country{China}
}
\affiliation{%
  \institution{The Hong Kong University of Science and Technology}
  \city{Hong Kong}
  \country{China}
}

\renewcommand{\shortauthors}{Rong Huang, Hai-Chuan Lin, Chuanzhang Chen, Kang Zhang, and Wei Zeng }

\begin{abstract}
Landscape renderings are realistic images of landscape sites, allowing stakeholders to perceive better and evaluate design ideas.
While recent advances in \gls{generative artificial intelligence} enable automated generation of landscape renderings, the \gls{endtoend} methods are not compatible with common design processes, leading to insufficient alignment with design idealizations and limited cohesion of iterative landscape design.
Informed by a formative study for comprehending design requirements, we present \emph{PlantoGraphy}, an iterative design system that allows for interactive configuration of GAI models to accommodate human-centered design practice.
A two-stage pipeline is incorporated:
first, \emph{concretization} module transforms conceptual ideas into concrete scene layouts with a domain-oriented large language model;
and second, \emph{illustration} module converts scene layouts into realistic landscape renderings using a \gls{finetune} \gls{lora}.
\emph{PlantoGraphy} has undergone a series of performance evaluations and user studies, demonstrating its effectiveness in landscape rendering generation and the high recognition of its interactive functionality.

\end{abstract}




\maketitle

\section{Introduction}\label{sec:introduction}

\begin{figure}[t]
	\centering
	\includegraphics[width=0.85\linewidth]{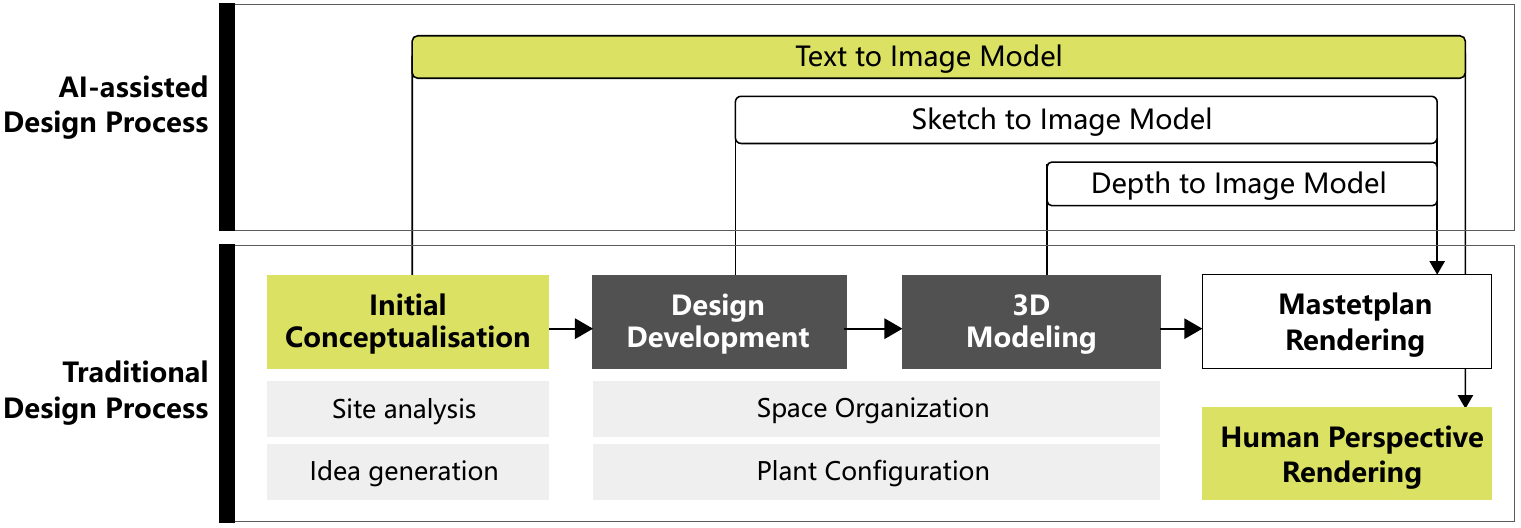}
	\vspace{-5pt}
	\caption{Comparison of traditional and AI-aided  design process for landscape rendering. Various AIGC models can be applied to different stages, yet these methods operate in an end-to-end manner without integrating iterative design.}
	\vspace{-3mm}
	\label{fig:design_process}
\end{figure}

While Goethe refers to architecture as frozen music, Filor draws an analogy between landscape design and ballet, which aims at discovering how the elements of nature can be recombined responsive to both planned and unforeseen uses on a daily and seasonal timescale~\cite{filor1994nature}.
The traditional landscape design process relies on the creativity and expertise of designers to create useful, comfortable and attractive spaces~\cite{boults2010illustrated}, which can be divided into four main stages: a) initial conceptualization, b) design development, c) 3D modeling, d) rendering, as demonstrated in Fig.~\ref{fig:design_process}(bottom).
Specifically, designers develop a preliminary concept taking into account site character, user acquirement and design vision, then refine the design concept into a detailed plan that includes specific design elements, such as plant selection and configuration.
The process outputs human-perspective landscape renderings that depict realistic images of a planned site with all plants, allowing stakeholders to evaluate the visual quality of landscape designs.

With the advances in generative artificial intelligence (GAI), AI-based image generation tools have emerged to facilitate the landscape design process.
\gls{pre-trained model}, such as Stable Diffusion~\cite{rombach2022high} and DALL-E-2~\cite{ramesh2022hierarchical}, can take descriptions for a garden like ``\emph{a realistic picture of a landscape design with trees, including a dogwood with pink flowers, flowering plants such as white tulips and daisies}'', and produce corresponding landscape rendering.
Due to the flexibility, GAI tools can play varying roles at different stages of the design process, as illustrated in Fig.~\ref{fig:design_process}(top).
Within the context of human-perspective landscape rendering, AI is acting as co-creators with the designers, refining the abstract concept and generating human perspective renderings through text-to-image models~\cite{rombach2022high,ramesh2022hierarchical}.

However, existing GAI tools primarily work in an end-to-end manner, \rev{lacking the flexibility necessary for designers to incorporate common design practices that iteratively refine the outputs.}
In particular, we identify the following limitations by existing \textcolor{blue}{AI-supported designing process from literature review (Sect.~\ref{sec:related_wotk}) and a formative study with landscape designers (Sect.~\ref{sec:overview}).}
\rev{1) \textit{Absence of interactive functionality for iterative design.}}
	Landscape design is an iterative process shaped by both concept and form, aiming at finding the approaches that best respond to expectations~\cite{filor1994nature}.
	Existing end-to-end approaches focus on design control within text editing, along with in-coherent results in multiple generations, causing absence of interactive functionality for iterative design.
\rev{2) \textit{Insufficient sensitivity to design arrangement of elements.}}
	Landscape designers are tasked with selecting plant species and determining the optimal combination expressed using directional words and commonly used plant words, which may not be accurately interpreted by text-to-image models.
	For example, existing \textcolor{blue}{\gls{artificial intelligence generated content}} tools are not compatible with descriptions like ``\emph{the daisy is located below the dogwood, and the white tulip is positioned to the right of the daisy.}''

To address these requirements \rev{from landscape designers}, we present \tool, an intelligent system with interactive functions for iterative refinement and an enhanced comprehension of landscape scenario descriptions.
As illustrated in Fig.~\ref{figure:framework}, \tool has two main modules: 
1) \emph{Concretization module} (Sect.~\ref{subsec:text2layout}) that transforms idea descriptions to scenario layout. 
The module introduces the concept of the scene graph, which represents landscape design as a structured, semantic description of the objects and their relationships within a scene.
Employing a dataset comprising scene descriptions, scene graphs, and layouts, this module utilizes a \textcolor{blue}{\gls{Large Language Model}} to transform the designers' descriptions into layouts with the guidance of scene graphs, which provide more interpretable and controllable information about the landscape scene for guiding the rendering generation task.
2) \emph{Illustration module} (Sect.~\ref{subsec:layout2SD}) that generates layout-guided landscape rendering.
We harvest a vegetation dataset commonly used in landscape design, with the participation of expert designers.
Utilizing this dataset, the \textcolor{blue}{\gls{latent diffusion model}}~\cite{rombach2022high,li2023gligen} is fine-tuned through a Low-rank Adaptation (LoRA) model~\cite{hu2021lora}, which enhances the model's ability to generate specific plants within the landscape scenario with high precision and accuracy. 
\tool also incorporates a web-based user interface (Sect.~\ref{subsec:user_interface}) that enables users to easily generate landscape design renderings and iteratively refine them by interacting with the graph and layout components. 

\textcolor{blue}{
A thorough assessment of \tool has been conducted from various perspectives.
For concretization module assessment (Sect.~\ref{subsec:experiment1}), two subjective experiments with quantitative metrics pertaining to layout were conducted, highlighting the essential role of incoperating scene graph in layout prediction and injecting domain knowledge using reasoning methods.
For illustration module evaluation (Sect.~\ref{subsec:experiment2}), three experiments in both objective and subjective perspectives were carried out, with results demonstrating the effectiveness of \tool in improving coherence in the results of multiple generations and comprehending designers' intents.
To further evaluate the proposed AI-assisted landscape design process, we conducted a within-subjects study (Sect.~\ref{sec:expert_interview}) that involves creating landscape renderings using \tool and traditional design software separately.
The results suggest that \tool holds varying degrees of value at different design stages. 
The findings also highlight human-centered AI assistance in design, emphasizing the importance of supporting iterative refinement of designs at each stage of the design process.
}

The major contributions and novel aspects of this work include:

\begin{figure}[t]
	\centering
	\includegraphics[width=0.99\linewidth]{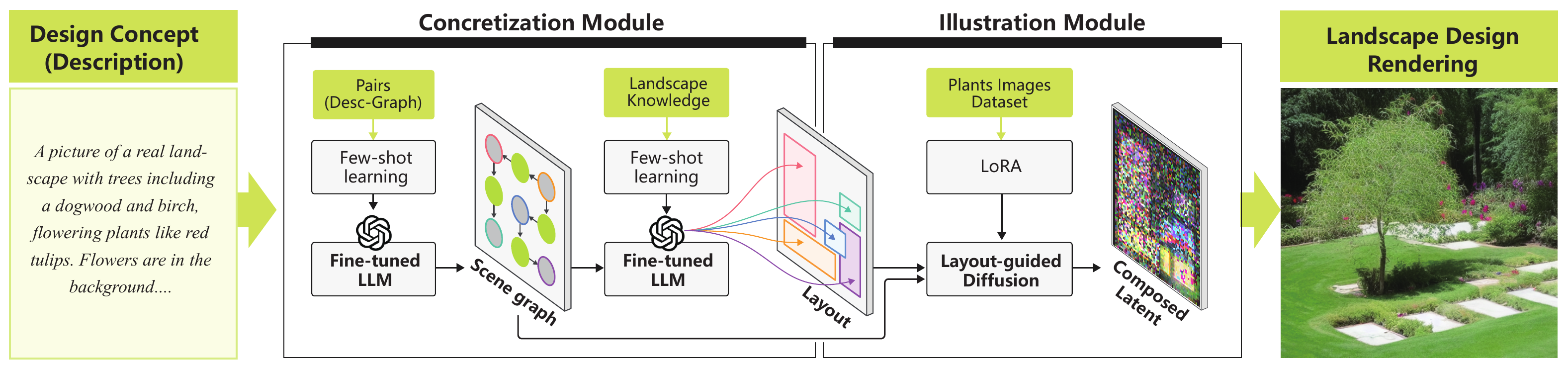}
	\vspace{-3mm}
	\caption{Overview of the system workflow. \tool incorporates a two-stage framework to transform design concepts described in textual content into realistic images of landscape rendering. First, the \emph{concentization} module leverages a graph-enhanced LLM to transform design descriptions into layouts, using scene graphs to improve the comprehension of user input. Next, the \emph{illustration} module employs a fine-tuned LoRA model to generate realistic landscape renderings based on the layout.}
 \vspace{-5mm}
	\label{figure:framework}
\end{figure}

\begin{itemize}[itemsep=0.1em,parsep=0.1em]
	\item \textbf{Framework}. 
	We propose a novel framework that incorporates an iterative design process into the AI-assisted designing pipeline for landscape rendering. 
	This framework maximizes control over the coherence in multiple generations of results, and addresses the challenge of the generative model's insensitivity to commonly used orientation words and plant terms in landscape design scene descriptions.
	\item \textbf{System}.
	We introduce an interactive system, namely \tool, to support the landscape rendering generation and iterative refinement.
	\textcolor{blue}{The system is built upon a custom dataset created in collaboration with expert designers that comprises realistic plants rendering images.}
	\tool speeds up the iteration by allowing designers to set constraints through interaction, promoting the output design to be more consistent with their expectations and increasing the involvement of landscape designers in the AI-assisted designing process.
\end{itemize}

\section{Related work}\label{sec:related_wotk}

\subsection{\textcolor{blue}{Generative AI for Creative Design}}
\label{subsec:AI-supported design process}
Generative models are capable of generating new data points that are similar to the training data set~\cite{Foster2023Generative}.
\rev{
The recent surge in GAI has ignited growing interest across various creative design fields, including fashion design~\cite{wu2023styleme, Lee_2023_fashioning}, UI design~\cite{kim2022stylette}, and visualization design~\cite{xiao2023let}.
GAI serves to either automate specific tasks or to foster the exploration of creative ideas~\cite{ko2023large} during the creation phase among various design processes~\cite{shi2023understanding, goldschmidt2014linkography}.}
Recently, there has been a growing interest in diffusion models~\cite{sohl2015deep, ho2020denoising}
\rev{among designers and artists}.
These include text-to-image (T2I) diffusion models~\cite{nichol2021glide,rombach2022high,ramesh2022hierarchical} that leverage textual \textcolor{blue}{\gls{prompts}} to direct the image generation process, and other conditional generative models like ControlNet~\cite{zhang2023adding}, T2I adapter~\cite{mou2023t2i} and GLIGEN~\cite{li2023gligen}, which allow more fine-grained input conditions.
\rev{While the remarkable efficiency and generation ability have been shown in technique manner, some limitations impede the usability of GAI in practical way of design.
The first and foremost reason is that, most diffusion models are realized in an end-to-end manner with input text prompts and output images, which overlook the “exploration” stage as one of the most important parts in creative process~\cite{inie2023designing}.
In addition, the lack of controllability cannot support stylistic consistency outputs, which is crucial for iterative design in practical scenarios~\cite{vimpari2023adapt}.}

\rev{This study aims to develop an AI-assisted design system for landscape rendering.
Three specific challenges are addressed:} integrating the design process into the model, obtaining suitable training datasets for this specific context, and exerting control over factors such as layout, plant positioning, and sizing.
While LLM-grounded diffusion~\cite{lian2023llm}, along with some other models~\cite{epstein2023diffusion, zheng2023layoutdiffusion}, claim to offer control over the shape, position, and appearance of generated objects, our experiments have demonstrated that their outcomes \rev{are hard to  meet} expectations of domain experts.
Furthermore, these methods do not encompass the final stage of rendering realistic landscape views.
To bridge this gap, we contribute a LoRA finetuning model trained on a landscape dataset that we have curated ourselves.

\subsection{Reasoning in Large Language Model}
LLMs have demonstrated promising results across a range of downstream tasks~\cite{shanahan2022talking}.
The capabilities have been exemplified by the GPT series~\cite{brown2020language,openai2023gpt4}, revealing emergent abilities that become apparent as the scale of training reaches a certain threshold~\cite{wei2022emergent}.
While LLMs are originally designed for text-based tasks, researchers have also ventured into other domains like vision-language reasoning.
For example, visual GPT~\cite{wu2023visual} fuses ChatGPT with visual foundation models to tackle vision-language tasks in an interactive manner.
Nevertheless, one significant challenge impeding the practical application of LLMs is their limited ability for reasoning.
In-context learning (ICL)~\cite{dong2022survey} addresses this challenge by enabling LLMs to generate expected outputs when given input text.
Studies have showcased the effectiveness of LLMs in solving complex reasoning problems through ICL~\cite{wang2023large}. 
Chain-of-thought (CoT)~\cite{wei2022chain} approaches are introduced to bolster the reasoning capabilities in complex tasks by introducing intermediate reasoning steps that lead to the final output.
CoT can be used with ICL in two main ways:  \textcolor{blue}{\gls{fewshot}} CoT and \textcolor{blue}{\gls{zeroshot}} CoT~\cite{kojima2205large}.
Few-shot CoT applies the step-by-step reasoning in the form of $<input, output> \to <input, CoT, output>$, whilst Zero-shot CoT directly generates intermediate reasoning steps to derive the answers, as exemplified by the phrase \textit{“let’s think step by step”}. 
Building upon the CoT framework, several works such as Auto-CoT~\cite{zhang2022automatic}, Tree-of-Thought~\cite{yao2023tree}, multimodal-CoT~\cite{zhang2023multimodal}, and multilingual-CoT~\cite{shi2022language} have been proposed to elicit reasoning abilities of LLMs in various tasks. 

LLM-grounded diffusion \cite{lian2023llm} combines GPT and diffusion models to perform conditioned text-to-image tasks, which can be applied to landscape rendering tasks.
However, this model does not inherently \rev{support iterative design process that is required by landscape designers.
The central challenge revolves around finding an effective method to integrate design rationales in the design process, into LLM-based design.}
To address this limitation, we leverage the scene graph concept, to represent plants within a landscape as entities and their spatial relationships as edges.
Experimental results demonstrate that this innovative approach can produce precise and realistic landscape renderings based on designers' intentions, \rev{and more importantly, facilitate iterative design process}.

\subsection{Scene Graph}

Scene graph is a structured data model employed to represent spatial relationships and semantic information of objects within a scene~\cite{johnson2015image,johnson2018image}. 
The ability to intuitively depict spatial relations has made scene graphs an emerging topic in both computer vision and AI research~\cite{chang2021comprehensive}.
As a graph-based representation, scene graphs can be seamlessly integrated with deep-learning-based generative models.
With advancements in conditional image synthesis, techniques have evolved to render images by conditioning generative models on scene graphs, such as GANs~\cite{ashual2019specifying,wang2022interactive} and diffusion models~\cite{yang2022diffusion}.
These methods allow designers and artists to work in a more abstract and intuitive manner by manipulating objects and their relationships rather than directly editing pixels or vertices.

Notably, the concept of the scene graph is highly aligned with the abstract bubble maps commonly employed in design sketch~\cite{herbert1993architectural,do2005design}.
The integration of scene graphs with generative models has the potential to enhance reasoning capabilities and provide users with greater control.
Scene graphs have been effectively employed in image generation through the prediction of bounding boxes and segmentation masks~\cite{johnson2018image,ashual2019specifying}.
However, these methods primarily rely on GAN-based approaches, which can present challenges in achieving high-resolution results and recognition of the rendering orders.
We address these challenges with LoRA diffusion models for high resolution renderings and an instance-based latent composition for controlling the overlapping order of objects.
Conversely, this study seeks to explore the possibility of leveraging scene graphs in conjunction with LLMs and diffusion models.
This represents a new and inherently challenging avenue of research, \rev{particularly for creative design}.


\section{Design Study}\label{sec:overview}

This section presents a group interview with landscape design experts aimed at gaining insights into the design prerequisites for collaborative efforts between humans and AI in landscape rendering (Sect.~\ref{subsection:group interview}).
In light of the findings, we consolidate design goals to be achieved for \tool (Sect.~\ref{subsec: design goals}).

\subsection{Group Interview}\label{subsection:group interview}
We conducted online interviews with five landscape design professionals (2 designers in employment \emph{U1} and \emph{U2}, 2 landscape design students \emph{U3} and \emph{U4}, 1 landscape design researcher \emph{U5}).
All designers have more than three years of landscape design experience.
Each interview lasted 40-60 minutes.
Specifically, as participants \emph{U2} and \emph{U4} had no prior experience using AI tools, we provided an introduction to existing tools and basic instructions before interviews.
We designed a question outline mainly focusing on three topics based on the relevant literature review and asked each designer to answer questions according to their design experience with no specific limitation: 
1) the general workflow of landscape design,
2) collaboration with artificial intelligence,
and 3) the pros and cons of existing AI-aided landscape designing tools according to their using experience.

At the end of the interview, we summarized the core insights based on the designers' feedback as follows:

\noindent
\textbf{Landscape designing process.}
We first surveyed relevant literature and industrial standard documentation to summarize the traditional landscape design process and what contents are usually included.
With details supplemented by the designers' practice experience, the traditional landscape designing process mainly comprises four stages (Fig.~\ref{fig:design_process}(bottom)): 
First, designers determine the initial design concepts according to the information collected from site analysis and cultural study.
Second, they develop designs by organizing space and flow on the sketch of masterplan.
Third, designs will be modeled in professional 3D modeling software and refine the construction detail.
Finally, designers will render the 3D model into a real scene through rendering engines to test the visual experience from a human perspective.
It's worth noting that participants mentioned that in the practice project, 

\begin{tightcenter}
\emph{"...design requirements are always adjusted, the aforementioned process is cyclical rather than sequential, until an optimal fit between concept and approach is achieved, as refereed iterative design." - U1 \& U2}
\end{tightcenter}

\vspace{1.5mm}
\noindent
\textbf{Collaboration with AI.}
AI can be utilized to assist at different stages of landscape design, as illustrated in Fig.~\ref{fig:design_process}.
When the input is a scenario sketch or a depth map, the design is completed by designers while the AI serves as an acceleration tool in the drawing process. 
Another situation is that the AI acts as a co-creator for the landscape designer when the input is an initial conceptual description \rev{expressed in textual prompts}.
Participants \emph{U3} and \emph{U4} expressed a preference for this co-creative approach, which allows designers to impart their ideas by delivering diverse interpretations of the \rev{textual} concepts.
\rev{This aligns with common design practices where designers often receive vague and abstract descriptions from clients.}
However, \emph{U1} and \emph{U2} expressed concerns about the potential loss of control for designers when using \rev{graphical} tools to bypass essential steps in the landscape design process. 
They all agree that when designers are deprived of multi-dimensional control over the outcome, the meaning of design is lost.

\centerline{\emph{"I feel like I'm more of a user than a designer." - U1}}

\vspace{1.5mm}
\noindent
\textbf{Existing tools for landscape design.}
All participants agreed that current tools can provide designers with multiple solutions to inspire them, but they rely heavily on text input, which is not always accurately understood by the model, particularly with regard to orientation descriptions.
Besides, participants expressed a preference for more interactive functions of graphical interfaces \rev{complemented with textual prompts that convey} the logic of designers' thinking, allowing designers to express their design intent in multiple ways (\emph{U3}).
\emph{U1} and \emph{U2} noted that each time they refined the description, the outputs by existing models varied drastically, far away from designers' needs to iteratively refine the design to accommodate changing conditions.
Additionally, designers need to select plant species considering the environmental conditions and characteristics of species and determining the best location for each plant based on its environmental requirements and aesthetic considerations. 

\begin{tightcenter}
\emph{"Current model has limited plant species available resulting in incorrect plant species generation." - U1 \& U3 \& U4}
\end{tightcenter}

\subsection{Design Goals} \label{subsec: design goals}
\rev{
Our objective is to harness human-AI collaborations to support the creative design process for landscape renderings.
This process entails progressing from general concepts to detailed elements, with iterative adjustments and refinements made along the way~\cite{shi2023understanding}.
In an AI-assisted design approach, the goal is to minimize repetitive tasks while enhancing the designer's creativity, rather than replacing them~\cite{amershi2019guidelines, capel2023human}.
To achieve this, the system needs to enhance its ability to interact with designers, accepting their guidance and actively suggesting possible solutions, thus completing a feedback loop in the creation process~\cite{yang2020re, shi2023understanding, shi2023hci}.
In line with the HCI community's exploration of augmenting user control over models, enhancing designer involvement in the AI-supported creating stage can fully leverage specialized knowledge~\cite{shi2023hci}.
The \gls{graphical user interface (GUI)} plays a crucial role as a user-friendly and efficient interaction approach to accept inputs from designers.
This includes using buttons and sliders for inputting basic settings~\cite{dang2022ganslider}, text editors for expressing design concepts~\cite{cheng2020sequential, wu2022promptchainer}, and graphical control for modification~\cite{wang2023pointshopar, chung2022talebrush}.}

Based on the feedback gathered from the design study \rev{and existing design practices}, we formulate a set of design goals for our intelligent landscape design system.

\begin{itemize}
\item
\textbf{G1: Improvement of the result's graphical coherence to support iterative design.}
To meet the iterative modification needs of designers, our system should enhance the graphical consistency of multiple generating results with the exception of the modified portion when user input is micro-changed, \rev{\eg, slightly changing the plants' position and sizes}.

\item
\textbf{G2: Enhanced model's comprehension of landscape design descriptions.}
Our system should be equipped with a model capable of comprehending design descriptions in landscape scenarios, particularly orientation words that are commonly used by designers. Additionally, it should support accurate generation for expert commonly-used plant species.

\item
\textbf{G3: Interactive editing function for in-depth designer participation.}
Our system should offer multiple ways for landscape designers to express their design requirements.
In line with the thinking logic of designers, \rev{enabling textual prompts as input is necessary}.
A graphic-based editing function is a suitable addition, enabling designers to interact with the system in a more intuitive and visual manner.

\end{itemize}


\section{PlantoGraphy System}\label{sec:system_description}
This section introduces \tool, a novel system that supports the steerable generation of landscape rendering from scene description and facilitates iterative design by interactive functionality.
We summarize how the system design responds to the design goals in Sect.~\ref{subsec:system_overview}, followed by a detailed description of the interface design (Sect.~\ref{subsec:user_interface}).
Then we introduce the main modules of the system: a \emph{concretization} module (Sect.~\ref{subsec:text2layout}) and an \emph{illustration} module (Sect.~\ref{subsec:layout2SD}).

\subsection{{System Overview}}\label{subsec:system_overview}
\rev{\tool is crafted for experienced landscape designers proficient in generating landscape renderings using the system and iteratively refining the results according to design requirements through the interactive editing module.}
To accomplish the goal, we design a two-stage approach as demonstrated in Fig.~\ref{figure:framework}:
\begin{itemize}[itemsep=0.1em,parsep=0.1em]
\item \textbf{Concretization module.}
The module harnesses domain-specific LLMs to translate users' conceptual ideas, as depicted in textual descriptions, into concrete scene layouts.
An innovation introduced here is the incorporation of \emph{scene graph} as an intermediary link between text and layouts.
Scene graph representation enhances the LLM's ability to comprehend landscape design descriptions (\textbf{G2}) when compared to direct conversion.
Furthermore, the scene graph acts as a medium for users to customize their designs, complementing text-only user interactions to improve graphical coherence (\textbf{G1}) and enhance interactive editing (\textbf{G3}).
The module offers \textit{scene graph} and \textit{layout} visual representations, to facilitate the iterative idea expression in an intuitive and interactive manner.
In the backend, two LLM-powered generators guided by well-structured prompt templates support the transformation between text, scene graph, and layout.
These prompts incorporate both reasoning enhancement techniques and domain-specific knowledge relevant to landscape design.
Furthermore, interactive features enable designers to make adjustments intuitively by editing the graph and layout. 
This includes actions such as adding or removing nodes and resizing elements to accommodate evolving design needs.

\item \textbf{Illustration module.}
The module employs a LoRA finetuning model to convert scene layouts into realistic landscape renderings.
One of the significant challenges faced in this process is the scarcity of available training samples.
To address this challenge, we have curated a vegetation dataset commonly utilized in landscape design with the invaluable input of expert designers.
The dataset encompasses a wide range of plant types, along with detailed attributes like common plant composition patterns.
This makes the generated outputs exhibit greater graphical coherence with user descriptions, aligning with \textbf{G1}.
Furthermore, we tackle the issue of overlapping object order, by implementing an instance-based latent composition process.
This module offers full automation, thereby saving designers time in design development and 3D modeling and empowering them to visualize how their conceptual ideas translate into human-perspective views.

	
\end{itemize}


\begin{figure}[h]
	\centering
	\includegraphics[width=\linewidth]{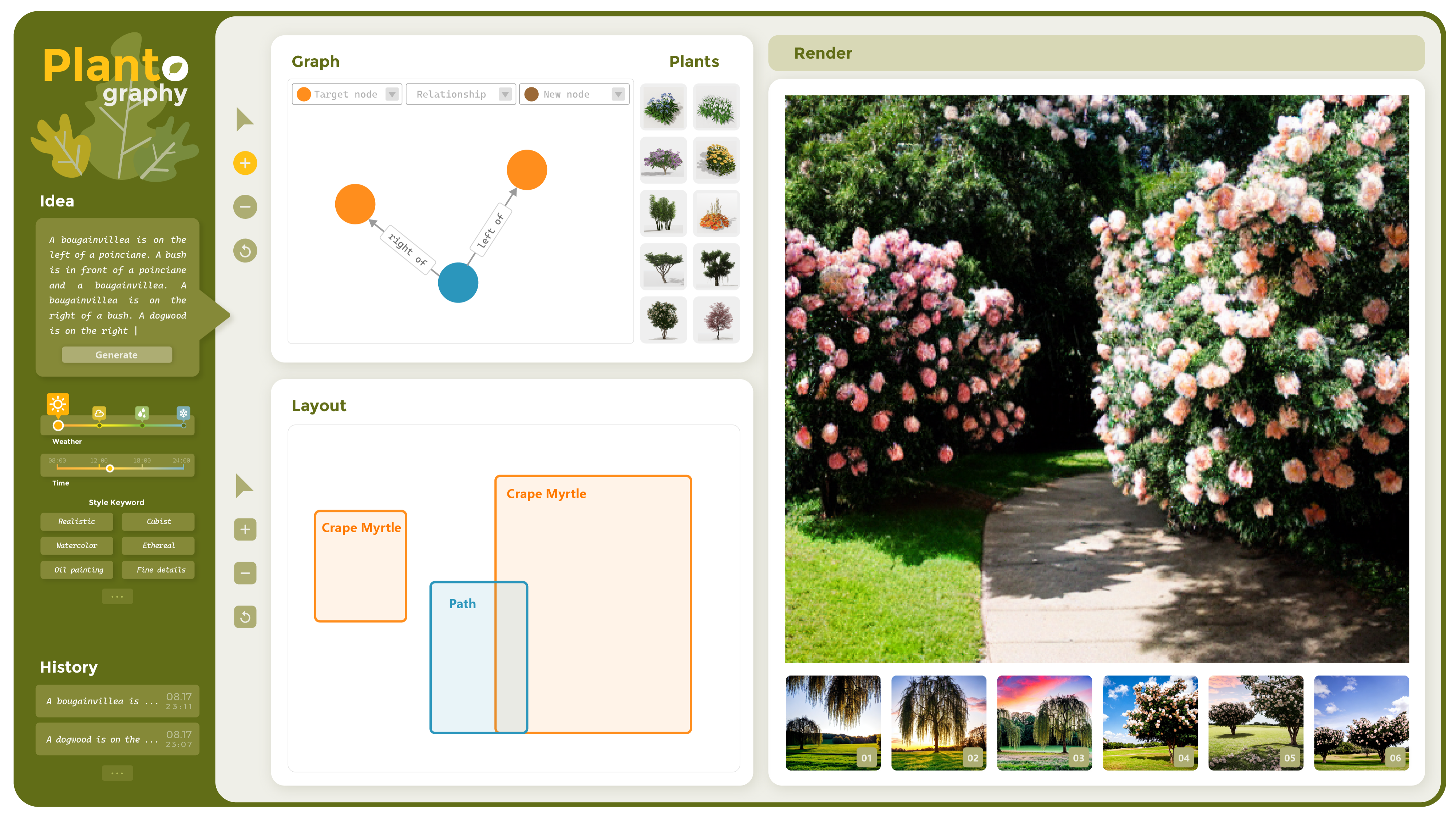}
 	\vspace{-7mm}
	\caption{Interactive visual interface for \tool. Users can input textual descriptions of the scene in the \emph{Text Panel} and customize the design by manipulating the scene graph in the \emph{Graph Panel} and updating the layout in the \emph{Layout Panel}. The rendering results are presented in the \emph{Rendering Panel}. }
	\label{figure:interface}
	 \vspace{-5mm}
\end{figure}

\subsection{User Interface}\label{subsec:user_interface}
\textcolor{blue}{An interactive interface is strategically crafted to bridge the gap between the rapid generation capabilities of current generative models and the controlled, iterative nature of traditional design processes. 
It empowers professional designers to harness the time-saving advantages of text-to-image technology while retaining as much control over the design development as traditional methods allow, ensuring a workflow that is both efficient and aligned with the natural progression of idea development. 
The interface comprises four panels depicted in Fig.~\ref{figure:interface}, allowing user to develop their idea in a procedural, controllable and interactive way: text, graph, layout, and rendering. }
\begin{itemize}[itemsep=0.1em,parsep=0.1em]

\item \rev{\textbf{Text Panel.}
The Text Panel initiates the design process.}
Designers can enter descriptive text, including plant species, quantities, and positional relationships between objects in the desired scene.
For example, \textit{"A realistic picture of a landscape design with trees, including a dogwood with pink flowers, flowering plants such as white tulips and daisies. The daisy is located below the dogwood, and the white tulip is positioned to the right of the daisy."}
\rev{To assist designers in avoiding the need to repeatedly input common constrain prompts, we offer commonly used options related to landscape design, including sliders for time and season, as well as rendering styles. Additionally, at the bottom of the panel, we provide a historical module to store previous design flows, allowing designers to iterate designs across different times and projects}

\item \rev{\textbf{Graph Panel.}
The Graph Panel organizes and visualizes the scene graph generated from designers' descriptions, while also enabling iterative adjustments to the scene graph.}
Each individual plant is represented by a node on the graph, and the oriented edge represents the relationship between nodes.
The label on the edge is a preposition indicating the orientation relationship between nodes.
Designers can create and manipulate nodes and edges to represent plants and their relationships.
A plant database panel is provided on the right side, allowing designers to directly drag the plant image as input.

\item \rev{\textbf{Layout Panel.}
The Layout Panel presents the inferred layout of the scene based on textual inputs and the scene graph, finalizing spatial details before transitioning to concrete landscape renderings.}
The panel displays each node in the graph as a box, with absolute positional
coordinates and accurate plant size, inferred from the concretization module.
Designers can adjust positions and sizes of elements to finalize the spatial layout.

\item \rev{\textbf{Rendering Panel.}
The Rendering Panel showcases the final landscape renderings generated by the system based on the intermediate scene graph and layout. 
The generated images by the iterative modifications are also stored at the history view in the bottom, which can be easily clicked on by the user to make comparisons.}

\end{itemize}


\subsection{Idea Concretization via Large Language Model}\label{subsec:text2layout}
\begin{figure}[t]
	\centering
	\includegraphics[width=0.95\linewidth]{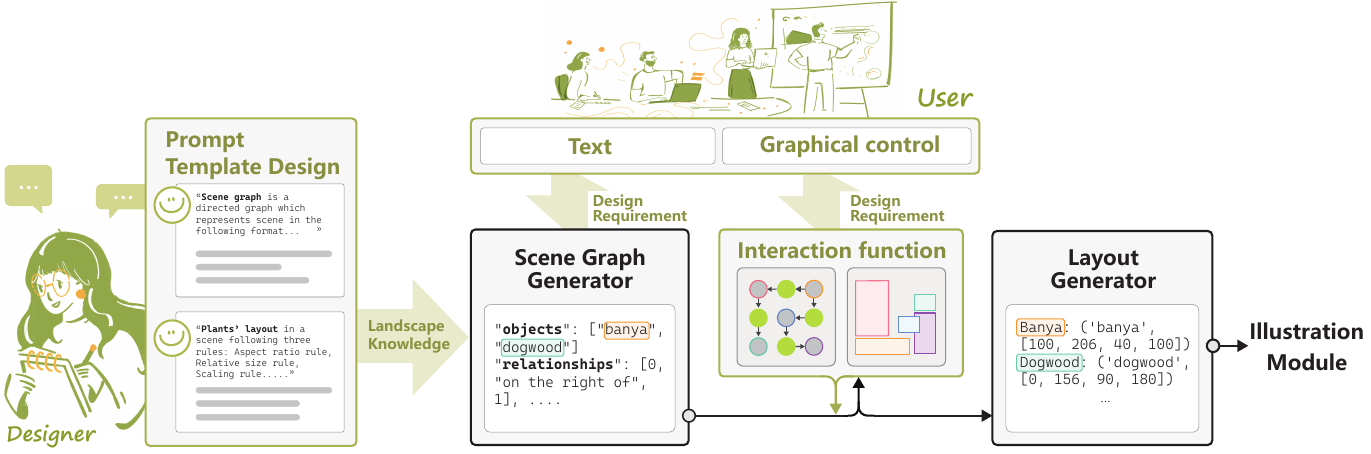}
	 \vspace{-2mm}
	\caption{Framework of idea concretization via expert-engaged interaction with LLMs for scene graph generator and layout generator.}
	\label{fig:LLM_human}
	 \vspace{-4mm}
\end{figure}

The \emph{concretization} module is designed to transform \rev{textual} scene descriptions into concrete layouts \rev{of design element \gls{boundingbox}}, serving as visual guidance and conditional input for the generative model in subsequent sections.
Drawing insights from the group interview (Sect.~\ref{subsection:group interview}), where designers \rev{expressed the expectations of supporting iterative design and improving the comprehension of design intentions}, we propose to integrate scene graph as an intermediate component in the process.
The scene graph aids in organizing design elements at an abstract level, avoiding early constraints imposed by concrete details and seamlessly bridging the transition from textual conceptual ideas to final layouts. 
As such, we develop \rev{two LLM generators: the first one transforms scene descriptions to a scene graph, and the second one transforms the scene graph to a layout separately}, through \textcolor{blue}{\gls{promptengineering}}. 
\rev{Specifically, the second generator is reasoning-intensive, requiring spatial and domain-specific reasoning in landscape design.
The LLM needs to infer the relative positions of plants in a two-dimensional space, deduce the size of each plant based on landscape knowledge, and refine the layout considering various constraints.}
The overall framework is illustrated in Fig. ~\ref{fig:LLM_human}. 

\begin{figure}[t]
	\centering
	\includegraphics[width=0.95\linewidth]{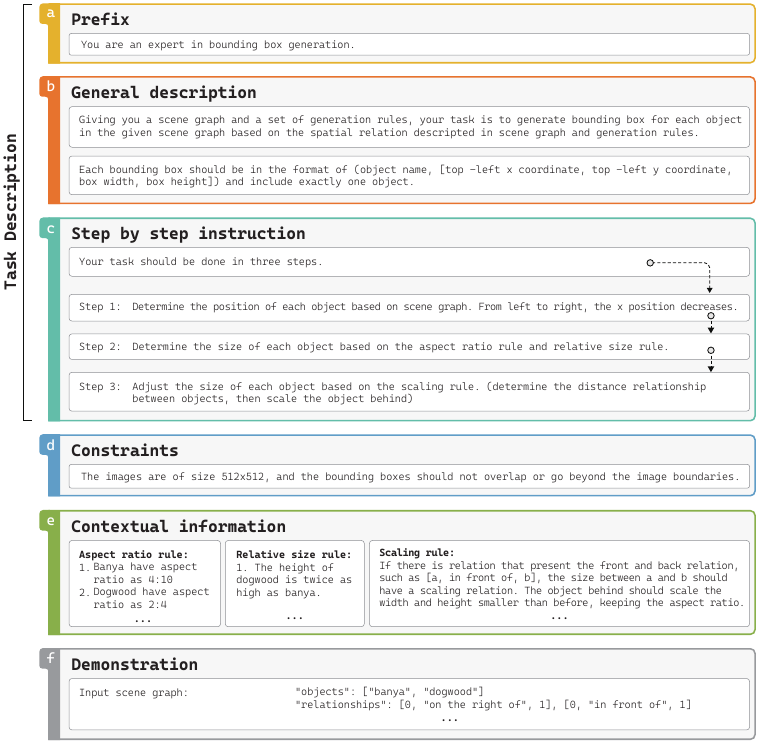}
	\vspace{-2mm}
	\caption{Prompt template for landscape scene layout generation. The template consists of four main components: task description, constraints, contextual information, and demonstrations. }
	\label{fig:prompt}
	\vspace{-4mm}
\end{figure}

Consequently, we have developed two distinct prompt templates, as shown in Fig.~\ref{fig:prompt}.
Both templates comprise four main components: the task description, constraints, contextual information, and demonstrations.
To equip the LLM's with domain-specific reasoning capabilities, the prompt for the graph-to-layout generator includes a landscape knowledge component as contextual information.
Below, we use the graph-to-layout generator as an example to illustrate each component. 
\rev{The complete prompt templates are shown in the Supplementary Material.}

\begin{itemize}
    \item 
\textit{Task Description:} The prompt commences with a prefix prompt (Fig.~\ref{fig:prompt} (a)) followed by a general description (Fig.~\ref{fig:prompt} (b)). 
The prefix prompt improves LLM's ability to perform specialized tasks with role conditioning.
The general description elucidates the input, output, and output formats.
We transform non-natural language entities, such as scene graphs and layouts, into a sequential format to enhance comprehensibility for the LLM.
The scene graph structure is linearized as a series of triples $<a, relation, b>$, where $a$ and $b$ represent nodes in the graph, and $relation$ corresponds to the edge connecting $a$ and $b$.
The layout is represented as $[object name, [x, y, width, height]]$.
To bolster reasoning capabilities during the generation process, we employ a chain-of-thought strategy as a step-by-step instruction (Fig.~\ref{fig:prompt} (c)).

\item
\textit{Constraints:} The constraints (Fig.~\ref{fig:prompt} (d)) define the boundaries within which the LLM is authorized to perform reasoning. Our prompts incorporate various types of constraints, such as restrictions in vision tasks and limitations on the number of generated elements in graphs.

\item
\textit{Contextual information:} While LLMs possess reasoning abilities, directly expecting them to reason with unfamiliar professional knowledge absent in the pre-trained model can be challenging.
Therefore, we introduce three key rules in landscape render generation for landscape design as contextual information for the LLM to reference during reasoning.
These rules encompass the aspect ratio of each plant, the relative sizes between plants, and the scaling effect for perspective relationship (Fig.~\ref{fig:prompt} (e)).

\item
\textit{Demonstration:} We provide \rev{5} question-answer examples to facilitate the LLM's reasoning ability and ensure alignment with the desired answering format through few-shot learning (Fig.~\ref{fig:prompt} (f)). 

\end{itemize}


\subsection{Landscape Illustration via Customized Diffusion Model}\label{subsec:layout2SD}

\begin{figure}[t]
	\centering
	\includegraphics[width=0.88\linewidth]{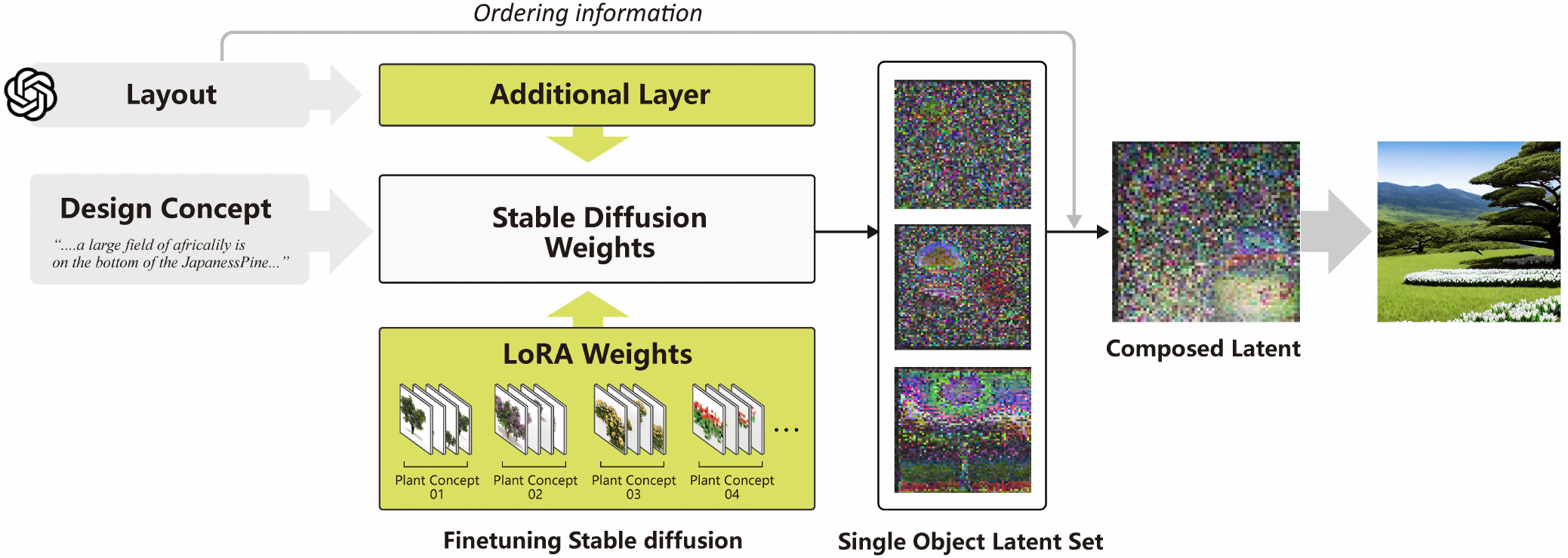}
	\vspace{-3mm}
	\caption{Framework of landscape illustration via layout-guided landscape rendering generation.}
	\label{figure:layout2img}
	\vspace{-4mm}
\end{figure}

The \emph{illustration} module takes the layout combined with textual prompts as inputs to produce realistic landscape renderings. 
\rev{The layout dictates the placement of plants, while the prompt specifies overall scene conditions like season and weather.}
To address the limitations of reliably generating specific plants and accurately deducing overlapping order faced by existing generative models, we have developed a more controllable framework, as illustrated in Fig.~\ref{figure:layout2img}.
This framework is composed of several components. 
First, we employ GLIGEN~\cite{li2023gligen} to act as the foundational model for open-set, layout-guided image generation.
We train LoRA models~\cite{hu2021lora} to fine-tune GLIGEN, enhancing its ability to generate domain-specific plants.
Additionally, an instance-based latent composition process is employed to manage the depth relationships among objects.
\textcolor{blue}{This integrated approach enhances the model's capacity to produce controllable and consistent landscape renderings, tailored to the requirements of professional landscape design.}

\subsubsection{Layout-guided diffusion model}
GLIGEN is used as its base model for high-quality, layout-guided image generation. 
GLIGEN enhances pre-trained models by integrating location inputs through a new trainable \textcolor{blue}{\gls{gated}} layer, preserving the original pre-trained weights. 
This enhancement allows the framework to effectively generate images based on both textual descriptions and layout information, while maintaining robust zero-shot capabilities.

\subsubsection{Model fine-tuning for better plant generation control}
To improve plant generation, we further perform fine-tuning of the base model using a self-curated dataset with the help of landscape designers, who provide their expertise and creativity in selecting the best plants and arrangements for each scene.
The dataset comprises \rev{hundreds of realistic rendering images in various scenes and includes 11 types of trees and 11 types of bushes}, with each plant type consisting of about 20 images.
The dataset is used to fine-tune LoRA, an effective fine-tuning method that updates only a small set of parameters in large pre-trained models.
We train LoRA models based on \textcolor{blue}{\gls{UNet}} and merge the LoRA weights with the frozen Stable Diffusion UNet part of GLIGEN.
Experiments show that the new types of plants introduced by our LoRA model can be correctly generated with layout conditions, even for never encountered concepts.

\begin{figure}[t]
	\centering
	\includegraphics[width=0.85\linewidth]{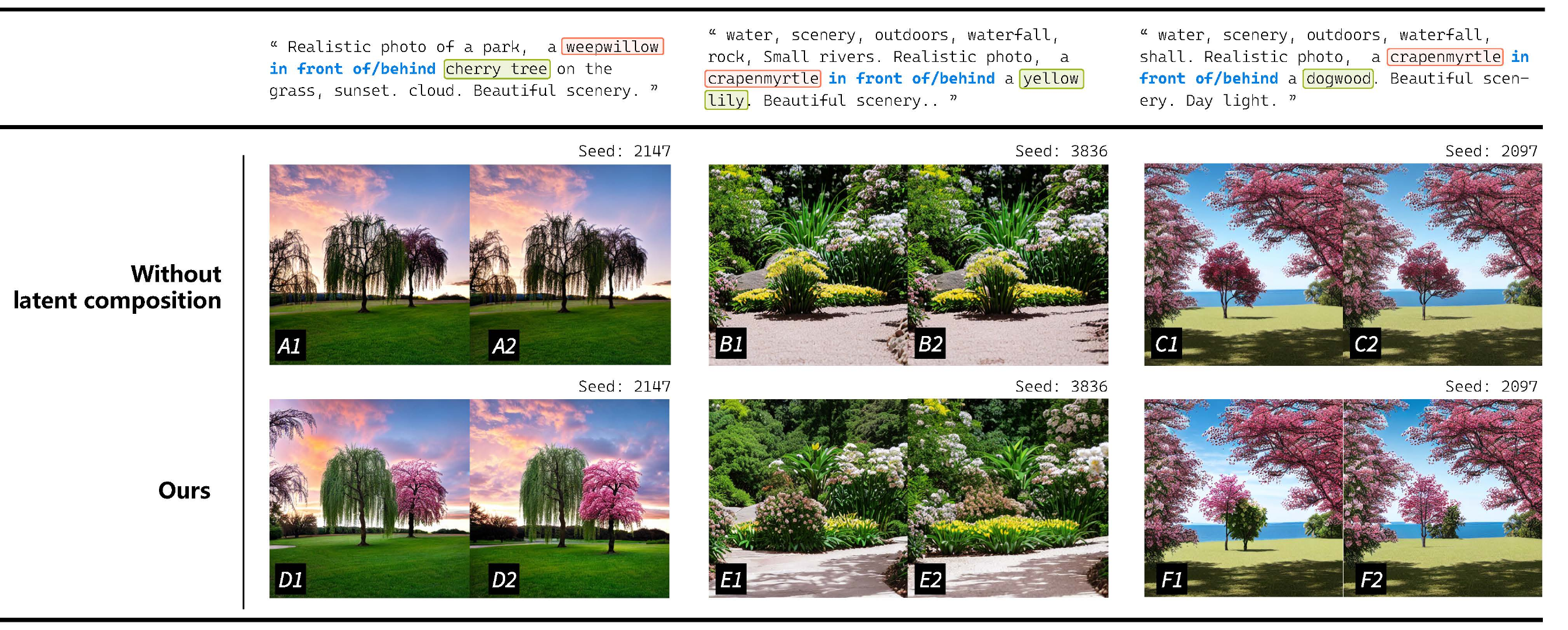}
	\vspace{-3mm}
	\caption{Results of with (bottom) and without (top) instance-based latent composition.}
	\label{figure:method_SD_ordering_results}
	\vspace{-2mm}
\end{figure}

\begin{figure}[t]
	\centering
	\includegraphics[width=0.85\linewidth]{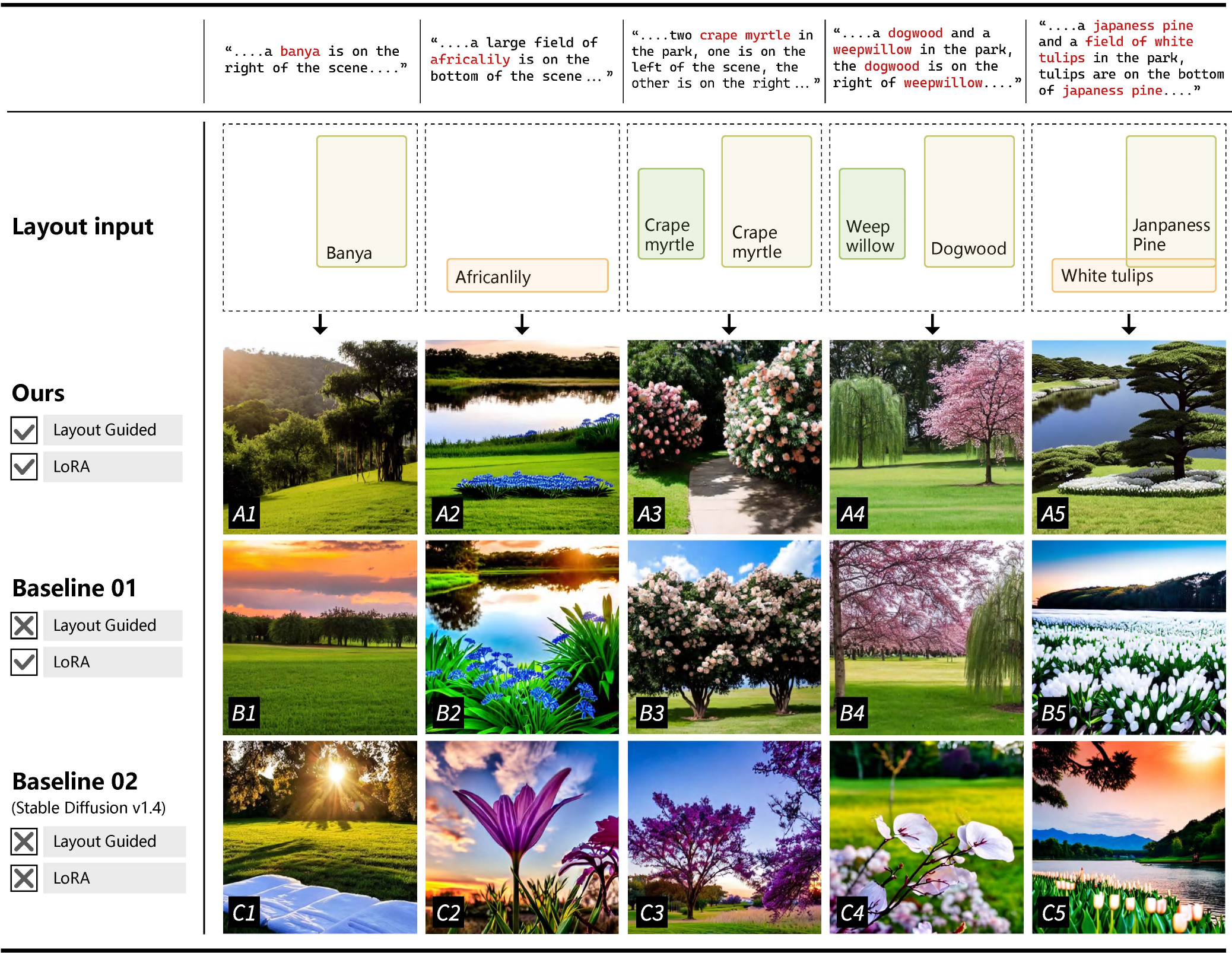}
	\vspace{-3mm}
	\caption{Comparison of generated results between our method to existing approaches.}\label{fig:model_result_comparision}
	\vspace{-5mm}
\end{figure}

\subsubsection{Instance-based latent composition}
Drawing inspiration from~\cite{lian2023llm}, we adopt an instance-based latent composition method, to address the limitation of GLIGEN's insensitivity to the front and back positions. 
The method treats each plant as a separate foreground instance, distinct from the background.
The generation process is outlined as follows:
\begin{enumerate}
    \item Utilize GLIGEN to generate images for each plant in the corresponding bounding box separately.
    \item Employ SAM~\cite{kirillov2023segment} to segment the largest object inside the bounding box for each plant. Typically, the largest object corresponds to the target plant due to the characteristics of GLIGEN. Store the segmented object masks.
    \item Utilize \textcolor{blue}{\gls{DDIM}} to retrieve the latent representation of each plant image.
    \item Determine the order of the plant objects within the scene graph, prioritizing them from the back to the front.
    \item Extract the latent within the segmented object mask for each latent representation of the plant image. Sequentially replace the extracted latent to the corresponding area (e.g., the segmented object mask) with a random Gaussian noise. The latent of the foremost plant object will be the last one to overlap. After this replacing process, the resulting latent representation is referred to as composed latent.
    \item Utilize GLIGEN once again to generate the landscape with the composed latent as initial latent. During the reference, all the areas containing plants will be frozen for certain steps.
\end{enumerate}
\rev{Fig.~\ref{figure:method_SD_ordering_results} compares the results by the models with and without instance-based latent composition, using the same prompts and seeds.}
The results indicate that the landscapes by ours adhere to appropriate foreground and background relations, as guided by the scene graph and the front-to-back order of the plant objects.
Unlike simply overlaying objects in the RGB pixel space, the overlapping in latent space ensures greater coherence with several inference steps.
\rev{Fig.~\ref{fig:model_result_comparision} presents generated landscapes based on the input prompts and layout, by our method (row 3) and two baselines (rows 4 \& 5).
The results demonstrate the necessity of incorporating layout guidance and model fine-tuning in our framework.}

\section{\lowercase{\color{blue}}{Experiments for Module evaluation}}\label{sec:evaluation}

\textcolor{blue}{
We conduct experiments to evaluate the performance of the proposed modules.}
\textcolor{blue}{1) Concretization module evaluation (Sec.~\ref{subsec:experiment1}):}
Three-dimensional quantitative metrics 
are employed to assess the accuracy of the LLM-powered generator in transforming text into the scene layout.
\textcolor{blue}{2) Illustration module evaluation (Sec.~\ref{subsec:experiment2}):}
We utilize structural similarity index metric (SSIM)~\cite{hore2010image} and conduct a controlled within-subjects user study, to evaluate the graphical coherence of the results generated during iterative design (Sect.~\ref{sssec:graphical_coherence}).
We further conduct a user study to verify whether our method succeeds in enhancing the models' comprehension of plant-related knowledge (Sect.~\ref{sssec:description_comprehension}).

\subsection{\textcolor{blue}{Experiment 1: Concretization module evaluation}}\label{subsec:experiment1}

\textcolor{blue}{\textbf{Metrics.}}
Following~\cite{lian2023llm}, we examine \rev{the} performance of the \textit{concretization} module for layout prediction from three dimensions: \emph{object-attribute assignment, spatial reasoning}, and \emph{perspective relational reasoning}, as follows:

\begin{itemize}
\item
\emph{Object-attribute assignment} is evaluated by two metrics:
First, \textit{correctness of aspect ratio}~\cite{zheng2021enhancing} measures the correctness of aspect ratio of a single plant, to assess object distortion in the generated renderings;
and second, \textit{correctness of relative areas} measures the correctness of relative size between plants, to assess visual coherence, balance, and logical structure of the plants.
\rev{Specifically, to measure \textit{correctness of aspect ratio}, 
we calculate L1 error between the aspect ratio of a ground-truth plant (denoted as $AR_{gt}$) and the one of the generation (denoted as $AR_{gen}$). A single plant aspect ratio is correct if $\|AR_{gt}-AR_{gen}\|_1< \theta$, where $\theta$ is set to 0.05.
We randomly sample 100 scene layouts as ground truths, and corresponding generation results from the generation method. A generated scene layout have correct aspect ratio if all plants in the generated scene layout have correct aspect ratios as the ground truths.
The \textit{correctness of relative areas} is computed similarly.
}

\item
\emph{Spatial reasoning} refers to the comprehension of the relative positions of plants, which is evaluated by the metric \textit{correctness of relative positions}.
Specifically, we utilize six common relative positions of `left', `right', `top', `bottom', `behind', and `in front of'.
\rev{For the metric, we use the same 100 scene layout samples and measure the relative positions of all pairs of plants based on their bounding boxes, and check if the corresponding results by the generation method have the same relative position. 
Specifically, if bounding boxes of two plants overlap, we manually check their relative position and categorize the relation as either `behind' or `in front of'.
}

\item
\emph{Perspective relational reasoning} further considers the correctness of scaling size due to perspective effects, for which objects that are further away (from human perspective) should have smaller relative sizes.
\rev{Here, we first compute the ground-truth relative size ratio $SR_{gt} = size(A_{gt}) / size(B_{gt})$ between two plants $A_{gt}$ and $B_{gt}$ based on a pre-defined dictionary of expected sizes according to domain knowledge. 
Then, we compute the actual relative size ratio $SR_{gen} = size(A_{gen}) / size(B_{gen})$ from the two generated plants.
Finally, we check the consistency of z-order relation between plants $A$ and $B$ with the order relation between $SR_{gt}$ and $SR_{gen}$, namely $A$ in front of $B$ $\Leftrightarrow SR_{gt}<SR_{ge+n}$, or $A$ behind $B$ $\Leftrightarrow SR_{gt}>SR_{gen}$.
For the metric, we use the same 100 scene layout samples and corresponding generated results. A generated scene layout have correct perspective relation if all pairs of plants in the scene layout have correct perspective relation.
}

\end{itemize}
\rev{The success of scene layout generation depends on whether these aspects are correct in the generated layout. In the experiment, the higher number of successful results in each aspect, the better performance of the model.}

\noindent
\textcolor{blue}{\textbf{Conditions.}}
We examine the performance of our method in comparison in five conditions: 1) zero-shot GPT3.5 with landscape knowledge, 2) few-shot GPT3.5 with landscape knowledge, 3) zero-shot GPT4.0 with landscape knowledge, 4) zero-shot GPT4.0 without landscape knowledge, and 5) few-shot GPT4.0 with landscape knowledge.
To ensure a fair comparison, each condition uses basically the same prompt architecture in \textcolor{blue}{Fig.~\ref{fig:prompt}} while only adding/removing the demonstration or background knowledge for the testing.
For each condition, we generate 100 samples for measurement,
\rev{based on sample prompts generated through a template that takes random plants and relations as text description.}


\begin{table}[t] 
\small
	\caption{The quantitative comparison between ours to ablation methods.}
	\label{table:layout prediction}
	\setlength{\arrayrulewidth}{1.2pt}
	\vspace{-11pt} 
	\begin{tabular}{lcccc}
		\hline
		\multicolumn{1}{c}{\textbf{Methods}}                                         & \textbf{\begin{tabular}[c]{@{}c@{}}Correctness of \\ aspect ratio\end{tabular}} & \textbf{\begin{tabular}[c]{@{}c@{}}Correctness of \\ relative areas\end{tabular}} & \textbf{\begin{tabular}[c]{@{}c@{}}Correctness of \\ relative positions\end{tabular}} & \textbf{\begin{tabular}[c]{@{}c@{}}Application of\\ scaling rule\end{tabular}} \\ \hline
		Zero-shot GPT 3.5  (w/  domain knowledge) & 51  & 56  & 58 & 66                            \\
		Few-shot GPT 3.5  (w/  domain knowledge) & 100  & 100   & 91 & 92      \\
		Zero-shot GPT 4.0   (w/o domain knowledge)         & 0     & 48   & 79  & 70   \\
		Zero-shot GPT 4.0   (w/  domain knowledge) & 82   & 94    & 72  & 70             \\
		\textbf{Few-shot GPT 4.0 (w/  domain knowledge) }& \textbf{100}                                                                      & \textbf{100}                                                                        & \textbf{92}                                                                            & \textbf{93}                                                                     \\ \hline
	\end{tabular}
	 \vspace{-5mm}
\end{table}

\noindent
\textcolor{blue}{\textbf{Results.}}
Table ~\ref{table:layout prediction} presents the count of successful results in four aspects between different models and methods. 
First, it is observed that both GPT-4.0 and GPT-3.5 achieve high generation performance with only a few errors in the generated results. 
The gap between GPT-4.0 and GPT-3.5 is not prominent, possibly because our prompts enable the GPT models to reach a certain high level at which the differences become negligible. 
For the zero-shot method, there is a significant gap between GPT-4.0 and GPT-3.5, suggesting that GPT-4.0 has superior zero-shot capabilities.
Focusing on each aspect, we note that the zero-shot GPT-4.0, without landscape knowledge for reference, generates zero correct aspect ratios and has a low correctness rate for relative areas when relying solely on common sense. 
This highlights the effectiveness of injecting domain knowledge into our method. 
On the other hand, for position reasoning and perspective reasoning, which involve common sense reasoning, the zero-shot GPT-4.0 outperforms the zero-shot GPT-3.5. 
It is evident that the few-shot model achieves a significant improvement over its zero-shot counterpart.
In summary, the results demonstrate the effectiveness of reasoning methods, such as few-shot learning, which are incorporated into our approach and highlight the essential role of landscape knowledge for language models to perform domain-oriented reasoning.

\subsection{\textcolor{blue}{Experiment 2: Illustration module performance}}\label{subsec:experiment2}


\subsubsection{G1: Ensuring graphical coherence of the results generated during iterative design}\label{sssec:graphical_coherence}

\leavevmode \\
\textcolor{blue}{\underline{Quantitative evaluation.}}
\textcolor{blue}{We first conduct a quantitative experiment to assess the efficacy of incorporating layout as a condition within rendering generation to sustain graphical coherence by comparing the similarity of images generated in multiple iterations.}

\textcolor{blue}{\textbf{Baseline model.}}
\textcolor{blue}{For comparison, we create a baseline model by removing the layout-conditional module but keeping other model configurations including the same LoRA model.}
This ensures consistency in the knowledge level between the models and minimizes the risk of misinterpretations that could influence the experiment.

\textcolor{blue}{\textbf{Data preparation.}}
We develop the test dataset from two sources: randomly generated and expert generated, with considerations of two variables: plant type and plant number.
First, we randomly generate 4800 landscape renderings including 5 compositions of landscape scenes that cover 20 different plant combinations: Tree A \rev{(C1)}, Tree A+Tree A \rev{(C2)}, Tree A+Tree B \rev{(C3)}, Tree A+Shrub A \rev{(C4)}, Shrub A \rev{(C5)}.
Second, we recruit 7 designers with over three years of experience in landscape design.
They are instructed to experience the interactive functionality of our system and use it to generate landscape renderings, fully simulating actual adjustments according to their professional experience.
The generation process mentioned above employs the same prompt \textcolor{blue}{with the same seeds group}. 
Each combination utilizes the same prompt and object layout while generated images differ according to random seeds.

\textcolor{blue}{\textbf{Metric.}}
We employ SSIM~\cite{hore2010image} to evaluate the images' graphical coherence during iterative generation.
SSIM is a widely used metric for measuring the similarity between two images. 
It takes into account the structural information of the images instead of only the pixel values, which is more in line with human perception and suitable for the assessment of layout-guided generation models.
The similarity of the generated images within the groups can be quantified by comparing the mean value of the SSIM.
A greater SSIM score signifies a heightened coherence among the images generated under identical prompts, thereby reflecting the robustness of the model in iterative design.

\textcolor{blue}{\textbf{Results.}}
As shown in Table~\ref{table: SSIM}, our model consistently outperforms the baseline in all categories.
The result signifies the effectiveness of introducing layout control in landscape rendering generation, particularly in enhancing image structural coherence throughout the iteration.
\rev{Notable, the improvements are more significant in C2 \& C3.}
\rev{This may be because plant types in C2 and C3 are both two trees, typically occupy a larger area than other groups in the image, making the consistency provided by the layout guidance more prominent across test sets.}

\begin{table}[t]
	\caption{Mean SSIM of images generated by two models.}
	\label{table: SSIM}
	\vspace{-9pt}
	\setlength{\arrayrulewidth}{1.2pt} 
	\begin{tabular}{lcccccc}
		\hline
		& \multicolumn{1}{c}{\textbf{C1}} & \multicolumn{1}{c}{\textbf{C2}} & \multicolumn{1}{c}{\textbf{C3}} & \multicolumn{1}{c}{\textbf{C4}} & \multicolumn{1}{c}{\textbf{C5}} & \multicolumn{1}{c}{\textbf{All}} \\ \hline
		\multicolumn{1}{l|}{{Baseline}} & 0.0956                       & 0.1017                         & 0.1123                        & 0.1138                        & 0.1387                        & 0.1091                          \\
		\multicolumn{1}{l|}{\textbf{Ours}}    & \textbf{0.1201}                        & \textbf{0.1466 }                        & \textbf{0.1858}                        & \textbf{0.1520}                        & \textbf{0.1545}                        & \textbf{0.1499}                         \\ \hline
	\end{tabular}
	\vspace{-5pt}
\end{table}

\begin{figure}[t]
	\centering
	\includegraphics[width=0.65\linewidth]{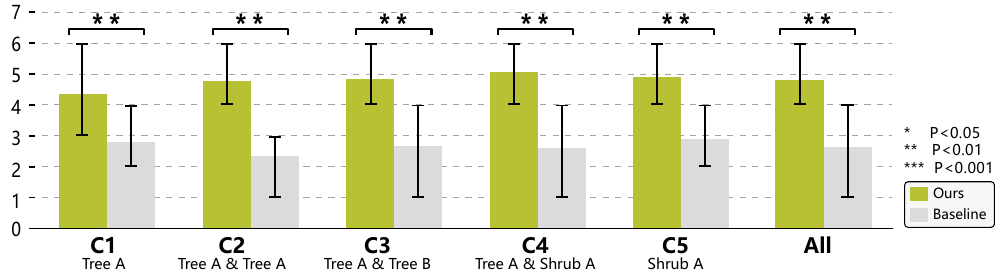}
	\vspace{-5pt}
	\caption{Results of the user study comparing the graphical coherence of the generated images within the groups.}
	\label{figure:evaluation_H1_userresults}
	\vspace{-5pt}
\end{figure}

\vspace{2mm}
\noindent
\textcolor{blue}{\underline{Subjective evaluation.}}
In landscape design, subjective perception of visual similarity can be affected by factors such as spatial ambiance, color and space harmony, which is not able to be evaluated using quantitative metrics.
\textcolor{blue}{As a complement from subjective perspective}, we further conduct a subjective user study to evaluate the graphical coherence considering human perception.

\textcolor{blue}{\textbf{Participants.}}
45 participants are recruited by posting recruitment messages on online social media.
Among them, 20 participants have design experience and the remaining 25 do not have.
The average age of the participants is 24.9 \textcolor{blue}{(2.22\% in [0,18], 53.33\% in (18,25], 35.56\% in (26,30], 8.89\% in (31,40])}.

\textcolor{blue}{\textbf{Procedure.}}
We randomly select 55 groups of generated images using each model with the same seed from the test set mentioned above. 
Participants are tasked to compare the graphical coherence of image groups, and \textcolor{blue}{give ratings form an 8-scale questionnaire}.
\textcolor{blue}{In the instructions given prior to the experiment, they are informed that 1) the questionnaire options represented perceptions ranging from "very dissimilar (0 scale)" to "very similar (7 scale)"; 2) the focus is on visual similarity, regardless of semantics; and 3) each question needs to be answered within 15 seconds.}


\textcolor{blue}{\textbf{Results.}}
\textcolor{blue}{No outlier is identified from the ratings.
We run the Friedman test on ratings for our approach and the baseline for each group.}
As demonstrated in Fig.~\ref{figure:evaluation_H1_userresults} (\textit{ALL}), our method outperforms the baseline model in terms of image coherence from a subjective perspective across all test groups. 
This is particularly evident in generating multiple objects, such as in groups \textit{C2, C3}  and \textit{C4}, \rev{where a comprehensive understanding of the relationships between these objects becomes crucial for analysis.}
For single object generation, our method presents a narrower performance gap compared to the baseline model, possibly because the baseline model possesses the ability to generate a single object with high robustness.
However, even in single object generation tasks, our method exhibits a significantly higher coherence compared to the baseline model. 
As the complexity of the generated images increases, the performance gap between our method and the baseline model widens. 
This indicates that our method is better suited for design tasks that require a higher level of coherence, especially when dealing with complex object relationships and compositions.

\subsubsection{G2: Ensuring the model's ability to comprehend landscape design descriptions}
\label{sssec:description_comprehension}

\leavevmode \\
We conduct a user study to evaluate the models' comprehension of designers' intentions, which is facilitated by the layout-based conditional input and LoRA module in integrating plant-related knowledge into the generation process.

\textcolor{blue}{\textbf{Baseline model.}}
To eliminate the impact of the layout condition on the generated results, we select our layout-guided model without the LoRA module as the baseline model.

\textcolor{blue}{\textbf{Data preparation.}}
For test set preparation, we randomly generate 50 images from each model covering 20 plant combinations the same as above.
The generated images are paired with corresponding generation seeds to create a 5-point Likert questionnaire.
10 landscape designers with more than three years of design experience are recruited to evaluate the alignment between design intent and images.

\begin{figure}[t]
	\centering
	\includegraphics[width=0.96\linewidth]{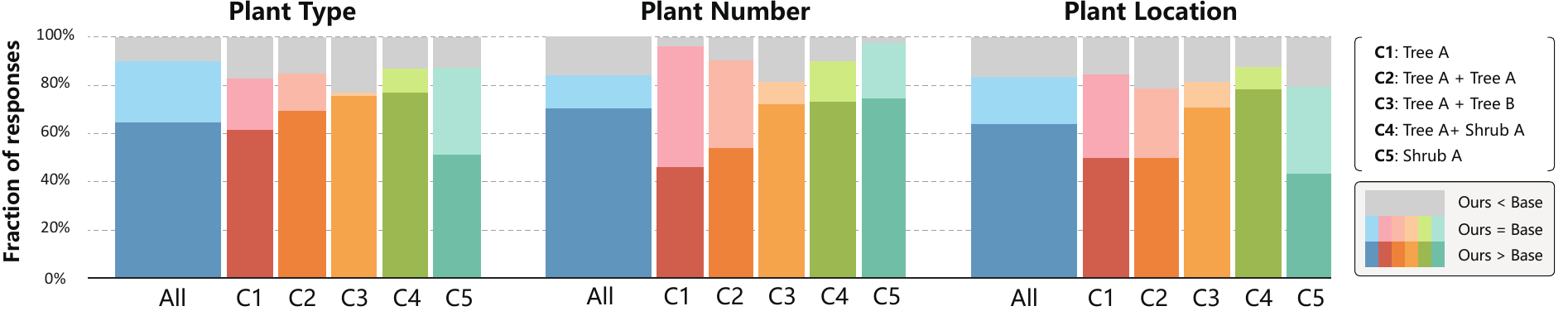}
	\vspace{-5pt}
	\caption{Result of the user study comparing the model's comprehension of landscape design descriptions.}
	\label{figure:evaluation_H2_results}
	\vspace{-5pt}
\end{figure}

\begin{figure}[t]
	\centering
	\includegraphics[width=0.99\linewidth]{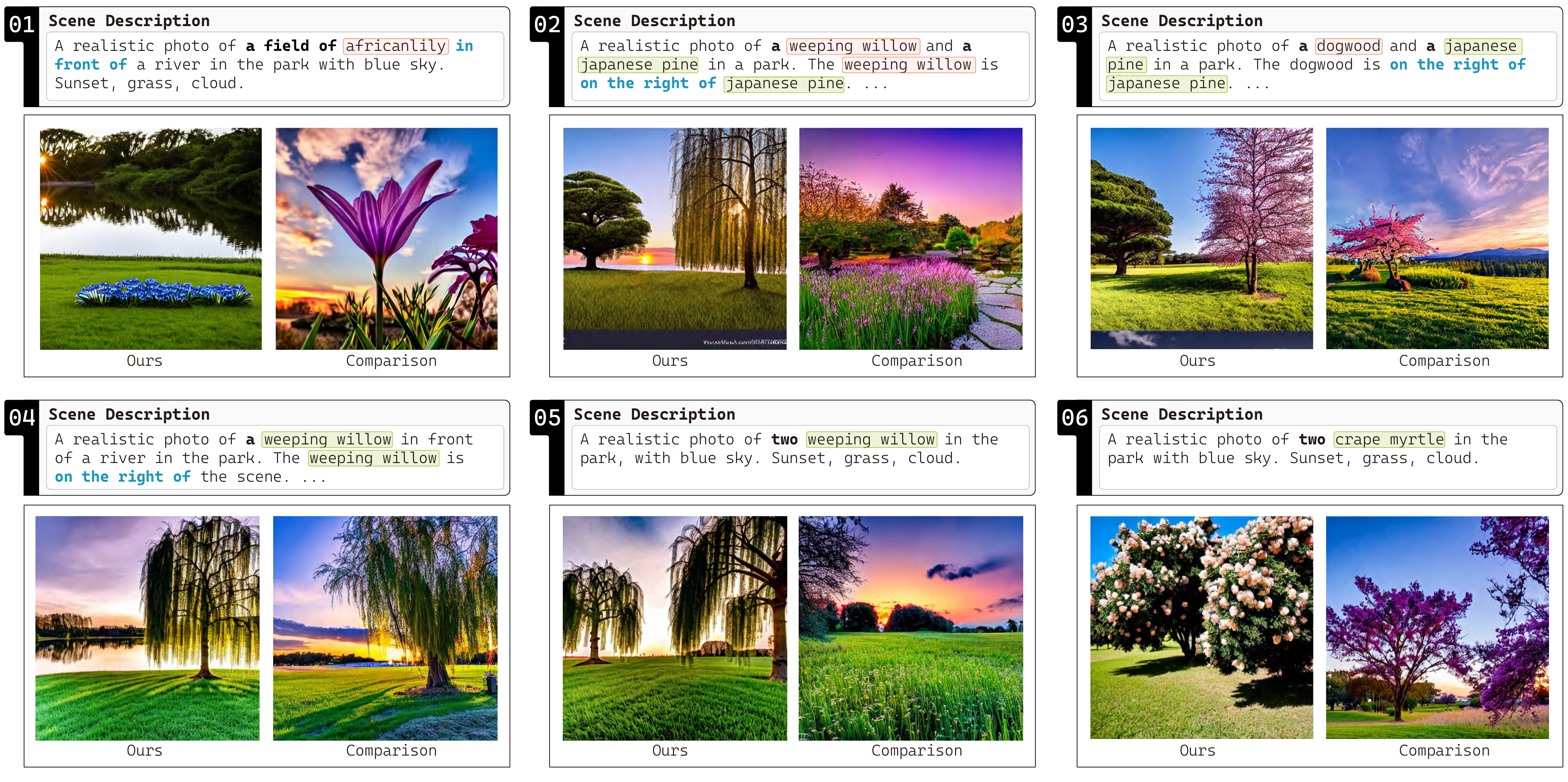}
	\vspace{-5pt}
	\caption{Cases with a significant difference between the evaluation of our model and the baseline model.}
	\label{figure:evaluation_H2_case}
	\vspace{-9pt}
\end{figure}

\textcolor{blue}{\textbf{Results.}}
As depicted in Fig.~\ref{figure:evaluation_H2_results}, our method exhibits a higher text-image alignment for plant type factors across all combinations, demonstrating that the LoRA module successfully integrates plant knowledge into the model and enhances its ability to accurately generate plant types.
In terms of plant type, groups \textit{C2, C3, C4} by our model all exhibit better performance than the baseline, indicating that our model is better at generating the appropriate plants in multi-object scenes.
As shown in Fig.~\ref{figure:evaluation_H2_case} (\textit{02}), the baseline model fails to generate the correct plants.
Specifically, \textit{C5} has the smallest gap between our model and the baseline model. 
This could be due to the fact that the shrubs in the baseline model overlap with the shrub species selected in our LoRA, resulting in both models performing well in this scenario. 

Unexpectedly, our method shows a better comprehension of plant numbers and plant locations, demonstrating that the layout-based conditional input can efficiently enhance the model's comprehension of plant numbers and locations. 
In contrast, in complex scenes (C3, C4), our model is able to better control the correct number and location of plants.
It is noteworthy that after adding plant species (C2/C3), the baseline model's ability to understand the number of plants decreased significantly, possibly due to the fact that pre-trained semantic information does not support such complex representations.
Fig.~\ref{figure:evaluation_H2_case} (\textit{04,05}) reveals that the baseline model can comprehend \textit{"a weep willow"} but fails to generate the correct scene for \textit{"two weeping willow"}.
In terms of plant location, our model still demonstrates superiority especially in complex scenes (\textit{C3, C4}).
As shown in Fig.~\ref{figure:evaluation_H2_case} (\textit{03}), the generated result omits the \textit{"Japanese pine"} and locates the \textit{"dogwood"} on the wrong side.
These underline the need to include layout control in models to increase graphical coherence and produce visually captivating images.

\section{User Study of \tool}\label{sec:expert_interview}
\textcolor{blue}{
	To further evaluate the effectiveness of \tool, we conducted a within-subjects study with 6 expert designers to compare the AI-supported landscape design process facilitated by \tool with the conventional design process using widely utilized industry software.
	}

\subsection{\textcolor{blue}{Participants}}\label{subsec: participants}
\textcolor{blue}{We recruited 6 experts (\textit{female}: 3 and \textit{male}: 3, \textit{Mean age} = 25.84) from social networks to participate in our study.
	To ensure a valid comparison, we selected participants who had experience using AI tools meanwhile had more than three years of experience with conventional design software.
	The demographics of all the participants are shown in Table~\ref{table:demographics}.
	The participants were invited to conduct the experiments offline.
	Given the complexity of the experiment tasks, we covered the travel expenses for all participants and compensated them at a rate of about \$13.5/hour.}

\begin{table}[h]\small
	\setlength{\arrayrulewidth}{1.2pt} \caption{Demographics of the participants in system evaluation.}
	\vspace{-5pt}
	\begin{tabular}{ccccc}
		\hline
		\multicolumn{1}{l}{\textbf{UID}} & \multicolumn{1}{l}{\textbf{Gender}} & \multicolumn{1}{l}{\textbf{Landscape design experience}} & \multicolumn{1}{l}{\textbf{Frequency on industry software}} & \multicolumn{1}{l}{\textbf{Frequency on AI tools}} \\ \hline
		E1                               & Female                              & 5 years                                                 & Very Frequently                                                    & Occasionally                                       \\
		E2                               & Female                               & 7 years                                                 & Frequently                                                    & Rarely                                              \\
		E3                               & Male                                & 7 years                                                  & Occasionally                                                & Frequently                                         \\
		E4                               & Male                              & 6 years                                                  & Very Frequently                                           & Occasionally                                              \\
		E5                               & Male                              & 5 years                                                  & Very Frequently                                              & Occasionally                                             \\
		E6                               & Female                              & 3 years                                                     & Frequently                                                     & Occasionally                                             \\ \hline
	\end{tabular}
	\vspace{-4mm}
	\label{table:demographics}
\end{table}

	\subsection{\rev{Experiment Setup}}
	\rev{For the user study, we devised a landscape design task simulating a real-world design creative process.
	The task was designed to assess participants' ability to utilize conventional designing tools \emph{vs.} \tool in creating landscape designs.
	The task involved three scenario descriptions as initial design requirements, as follows:
	\begin{itemize}
		\item Scenario 1: Wooded Path: A trail with tall trees on both sides. 
		\item Scenario 2: Lake and Cherry Trees: The path bordered by a lake on one side and a row of cherry trees on the other, with some flowering plants around the lake. 
		\item Scenario 3: Houses and Fields of Flowers: Small houses in the distance, with large fields of flowers nearby.
	\end{itemize}
	The participants were given the following task.
	\begin{tightcenter}
		\textit{ "You receive vague scenario descriptions from a client. \\Prepare a concrete landscape rendering for the next debriefing session, to help confirm the project requirements."}
	\end{tightcenter}
}

	\subsection{\rev{Experiment Procedures}}
	\rev{
	Prior to the study, the participants signed a consent form, agreeing to join the experiment, and allowing us to collect basic demographic information and record their behavioral data, including the frequency of operations on each function and the time taken to complete each task.
	Then, participants were instructed to go through the following steps:
}

	\begin{enumerate}[itemsep=0.1em,parsep=0.1em]

	\item \rev{Participants were briefly introduced to the background of the project and received an introductory about the usage of \tool interface for about 5 minutes. Then they were allowed to freely explore \tool for about 10 minutes.}

	\item \rev{Participants were asked to complete the design task using conventional landscape designing tools \emph{vs.} using \tool.
	For conventional design process, participants were provided with the flexibility to choose commonly used software that facilitates their design goals including modeling software (\eg, Rhino, Sketchup, C4D) and rendering tools (\eg, V-Ray, Lumion, Unreal Engine, Adobe Suite).} 

	\item \rev{Participants were asked to fill a post-study questionnarie with 5-point Likert scale questions focusing on 
		\emph{usability}, \emph{enjoyability} and \emph{effectiveness} ratings of each interactive panel and the whole \tool system.
	Additionally, participants were asked to answer four open-ended questions: }
		\begin{itemize}
		\item \rev{Does the interactive system improve the efficiency of collaborating with generative AI? If so, how?}
		\item \rev{What is the difference between GAI-supported design process and traditional design process?}
		\item \rev{What possible improvements can be made to the \tool system?}
		\end{itemize}

	\end{enumerate}
	\rev{To resolve bias by task familiarity, we randomized the order in which each participant using conventional tools or \tool.
	We did not impose any time constraints to encourage participants to focus on the iterative design and prioritize the quality of their results.
	The average time for completing the study was 103.5 minutes for each participant.}

\vspace{-3pt}
\begin{figure}[t]
	\centering
	\includegraphics[width=1\linewidth]{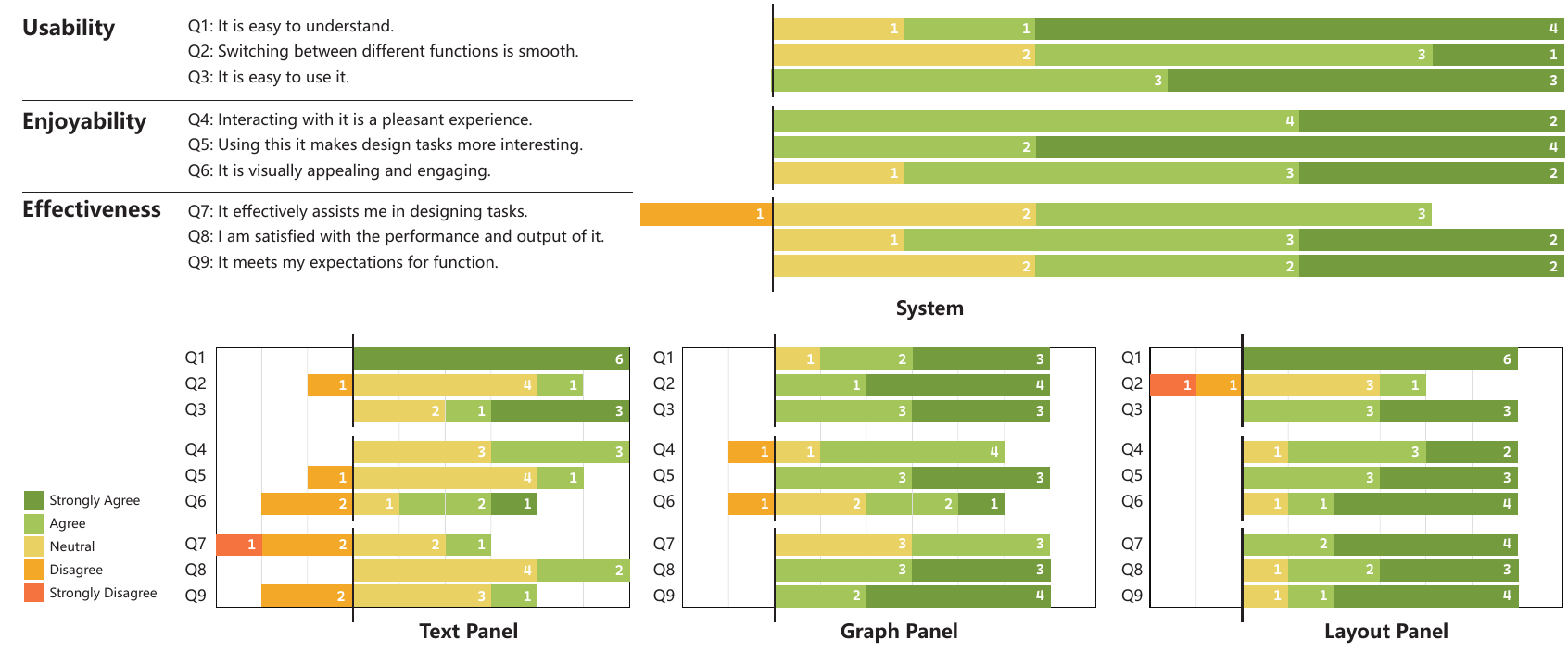}
	\vspace{-4mm}
	\caption{Participant ratings on the \tool system and the interactive panels in terms of \emph{usability} (Q1-Q3), \emph{enjoyability} (Q4-Q6), and \emph{effectiveness} (Q7-Q9).}
	\vspace{-3mm}
	\label{fig:perception}
\end{figure}	

\begin{figure}[t]
	\centering
	\includegraphics[width=0.85\linewidth]{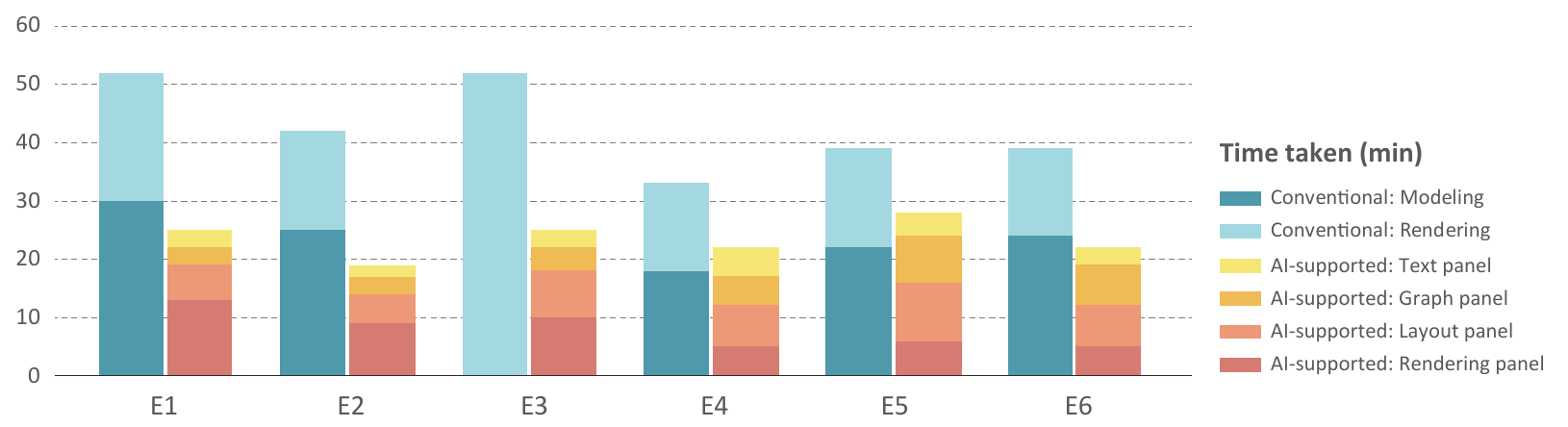}
	\vspace{-3mm}
	\caption{Time taken on design tasks using conventional and AI-supported landscape design process by each participant.}
	\vspace{-3mm}
	\label{fig:time taken}
\end{figure}

\subsection{\textcolor{blue}{Results}}
\rev{Below we first report the quantitative ratings and qualitative feedback from the participants on \emph{system design}, followed by a \emph{lessons learned} on AI-assisted design process.}

\subsubsection{\rev{System Design}}
\rev{User ratings regarding design of the overall system and interactive panels are depicted in Fig.~\ref{fig:perception}.}

\begin{itemize}	

\item \new{\textbf{Overall System.}
Participants generally expressed satisfaction with the overall system's \emph{usability} (Q1-Q3), \emph{enjoyability} (Q4-Q6), and \emph{effectiveness} (Q7-Q9).
Specificly, the majority of users rated the system with the highest score in easy to understand (Q1) and to use (Q3).
\textit{E4 \& E6} emphasized the importance of an intuitive interface for users engaging in design process.
Most users enjoy using the \tool system during creative process, with the highest score on making the design tasks more interesting (\textit{Q5}).
The participants unanimously lauded the system's interactive features, noting that "graphical editing panels align more effectively with typical design logic compared to conventional design process that requires several different softwares" (\textit{E3}).
Participant' response on the system effectiveness are relatively moderate.
Notably, we can see a concern on assistance in designing tasks while some users gave a low score (\textit{Q7}).
Participant (\textit{E6}) suggested that fine-tuning operations could be more intuitive if the layout were to be overlaid on the rendering panel.
Additionally, \emph{E1} observed that generating certain plant pairings, like weepwillow and banyan, can result in unusual edge formations in the plants.
More discussions are provided in the lessons learned below.
}

\item \new{\textbf{Text Panel.}
Participants showed a low level of interest on the Text Panel since the tasks were finalized and did not require any modifications.
	Interestingly, \emph{E3} raised a high point in enjoyable for text panel: \textit{"I like the weather and time slider that allow me to change the environment effect of generated image intuitively."}
	For improvement, participant suggested adding sketches as a more natural alternative for input, favoring their intuitive nature over traditional text-based methods (\textit{E6}).
	}

	\item \new{\textbf{Graph Panel.}
	The ratings on the Graph Panel are mostly positive.
	\textit{E2} found the graph representations and interactions especially beneficial, highlighting their potential to deepen understanding of complex structures.
	\textit{"Plant combinations are influenced by various factors, including local climatic conditions, applicable plants, and the relationships between plant populations. ... Graph can abstract these relationships and it is intuitive (E2)"}.
	\textit{E3 \& E6} suggested enhancements in the graph panel for smoother node addition and automated optimal node placement recommendations.
	}

\item \new{\textbf{Layout Panel.}
	Participants perceived the Layout Panel to be the most enjoyable and efficient among the three panels.
	The layout panel was predominantly utilized, overshadowing the frequency of the graph and text panels, aligning with the subjective scoring provided by users.
	\textit{E3} explained that \textit{"designers prefer to use graphic-oriented approaches conveying design concepts with stakeholders."}
	\textit{E5} also highlighted that "\emph{using a graphical editing function for minor modification is more effective for iterative design}".
	However, \emph{E4} gave the lowest score on switching between different functions (Q2), as the effect of layout modification was only visible on the output but not reflected on the other two panels. 
	}

\end{itemize}

\noindent
\new{\textbf{Time Analysis}.
Fig.~\ref{fig:time taken} illustrates the time spent by each participant on the tasks using conventional software and the \tool system.
Throughout the study, the traditional process demanded more time than AI-assisted ones.
The reduction in time is quite significant (mean: 20.17 minutes, max: 27 minutes, min: 11 minutes), showcasing the efficiency of \tool in streamlining design processes.
Notably, the time reductions are more significant by E1$\sim$E3 than those by E4$\sim$E6. 
The follow-up interviews revealed the reasons, as E1$\sim$E3 took longer times for conceptualization that solely relies on designers' knowledge and experience using conventional tools, whilst GAI can quickly give some prototype designs and help designers brainstorm.
Interestingly, E3 exclusively utilized Photoshop to merge online-retrieved plant images and create renderings.
However, this approach proved time-consuming, as substantial effort was required to manually adjust plant sizes and positions to accurately portray human-perspective views.}

\rev{We also observed distinct patterns in the temporal distribution of user engagement with the interactive panels in \tool during creative tasks (Fig.~\ref{fig:time taken}).
Here, time on the Rendering Panel pertains to the waiting durations for the model to generate renderings.
Undoubtedly, participants devoted the least amount of time to the Text Panel and the most time to the Layout Panel, underscoring their inclination for iterative designs.
}
 
\subsubsection{\rev{Lessons Learned}}
\new{Participants highly praised the system's ability to support them in iterative design. 
\tool provides opportunities for applying designers' expertise and collaborating with GAI, facilitating the development of distinctive and captivating designs.
However, the improvement is constrained by \tool's limited flexibility in accommodating more diverse and complex design requirements.
Participants attempted to incorporate additional plants that harmonized well with those in the given scenarios but encountered challenges, particularly when the number of plants exceeds six.
This is likely attributable to the restricted quantity of plants in the fine-tuning dataset, and the LLM's reasoning capability when dealing with intricate graphs.
}

\new{
Upon conducting a detailed analysis of the time reduction disparity between E1$\sim$E3 and E4$\sim$E6, we discovered that E1$\sim$E3 primarily focused on the initial phase of the design process, whereas E4$\sim$E6 shifted their attention to the more advanced and in-depth design phase.
E4$\sim$E6 noted that, "\emph{for the more advanced stage, the content and quality of the images generated better matched my graphic sketches}," necessitating additional iterative adjustments when using \tool.
The elongation of time in the traditional process is less pronounced, as the intermediate `3D model' outcome empowers designers with enhanced control for viewing and adjusting the design from various angles.}

\rev{This observation reveals a compelling aspect of AI integration in design processes:
as AI becomes more deeply embedded in the design workflow, its efficiency gains for designers seem to diminish.
This phenomenon may stem from the escalating complexity of decisions and creative inputs required at advanced stages, a realm where AI-assisted design tools may not yet exhibit the same efficacy as in initial stage tasks.
We anticipate that the progress of AI will enhance the effectiveness of AI-assisted design tools for more advanced tasks.
Importantly, this also implies a potential reevaluation of the role of AI: from a tool that expedites tasks to one that evolves to enhance, rather than overshadow, human creativity and expertise.
}
\section{Discussion}\label{sec:discussion}
\subsection{Findings}
	\subsubsection{Domain knowledge embedding}
	In our workflow, the end-to-end process of text-image generation is broken down into multiple steps, allowing domain knowledge to be introduced into the model development process in various forms: building prompt templates, translating experience into rules, etc.
	For example, the \emph{concretization} incorporates designers' experience and knowledge into the domain-oriented LLM through prompt templates.
	\rev{An advantage is that} we only need to construct a small number of datasets with experts and embedded the expertise into the model by means of few-shot learning, which is proved to be effective in Experiment 1 (Sect.~\ref{subsec:experiment1}).
	In this way, we utilize the powerful language processing and reasoning capabilities of LLMs to facilitate AI-assisted design.
	It requires less specifications of domain knowledge meanwhile retains high compatibility, which is friendly to non-AI practitioners.
	
	\subsubsection{Interaction \rev{design for generative AI}}
	Our system introduces scene graph and layout as two intermediate graphical representations in the text-image generation process. 
	These representations transform text-based scene descriptions into intuitive, easy-to-modify graphs from abstract and figurative perspectives. 
	In post-study interviews, users highlighted the benefit of multimodal inputs (\eg, text and graphical control) \rev{enabled by the GUI and widget interface}, allowing for more effortless and intuitive expression of design requirements. 
	This facilitates the accurate translation and communication of the designer's needs to the \rev{generative AI} models. 
	Positive user feedback on interaction modes suggests that designing functions tailored to the target population's thought processes and operational logic reduces the learning curve and facilitates smoother information transfer.
 	For example, designers in our study expressed familiarity with abstract graphs, aiding them in visualizing the system's potential for diverse design needs.
 	\rev{These findings align with the guidelines for crafting human-centered generative AI systems as outlined in~\cite{shi2023hci}, mirroring findings from recent HCI-centric GAI studies for various domains (\eg,~\cite{wu2023styleme, vimpari2023adapt}).}

\subsubsection{\rev{Control from designers to models}}
	\rev{Control from humans to models is regarded as a key dimension for human-centered generative AI systems~\cite{shi2023hci}.}
	In our approach, the use of layout as a conditional input enhances the graphical coherence of the generation model across iterations. 
	Users found this improvement to be highly practical for real-world design scenarios, indicating that the AI tool effectively integrates into the iterative design process and co-creates with designers. 
	Furthermore, controllable AI generation tools can empower designers to fully leverage their expertise and experience to stimulate creativity. 
	Throughout the design process, designers can freely explore and experiment with various creative solutions, with AI serving as an assistive tool to offer additional possibilities and inspiration. 
	In the future, AI generation tools incorporating \rev{multiple mediums of control} will become increasingly crucial. 
	Such tools will enable designers to achieve high-quality design outcomes more rapidly and enhance their work efficiency, all while capitalizing on the boundless creative potential offered by AI technology.
	\rev{However, it's essential to view AI tools as complementary, not substitutes, for professional landscape designers, as human expertise remains indispensable for ethical considerations, nuanced understanding, and client prioritization in landscape design.}

\begin{figure}[t]
	\centering
	\includegraphics[width=0.99\linewidth]{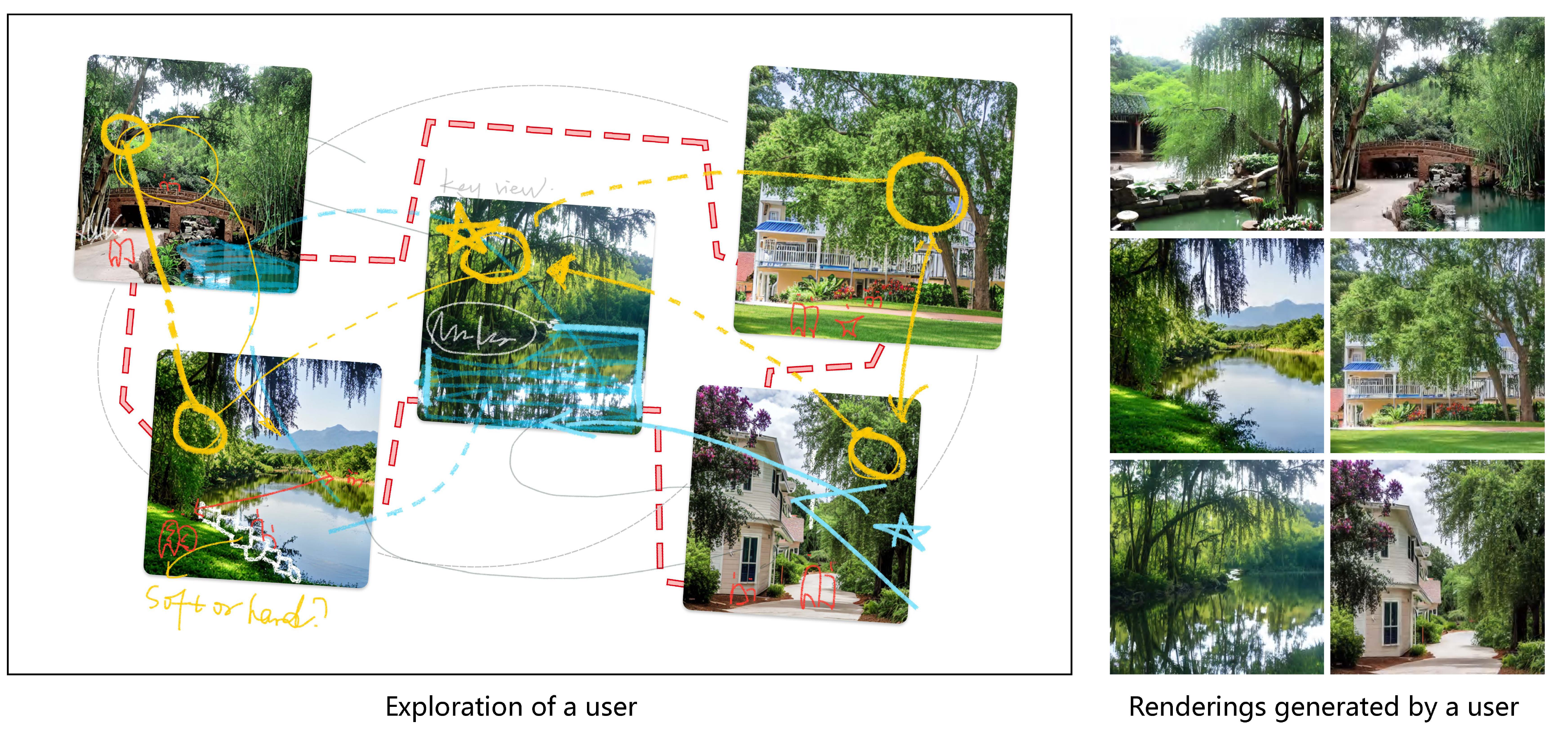}
	\vspace{-2mm}
	\caption{Narrative map generated by a participant with the help of \tool, illustrating his imagination of the human perspective key plots of the design site.}
	\vspace{-4mm}
	\label{fig:narrativemap}
\end{figure}

\subsubsection{AI-aided \rev{creativity}}
	\rev{Many studies have demonstrated the capability of GAI systems in fostering creativity, by controlling parameters like random seed or penalty to create randomness in the outputs~\cite{han_2022_initial,Louie_2020_novice}.
	\tool also allow users to control the inputs to create diverse landscape renderings.
	Moreover, our system further introduces creativity by introducing multiple steps when utilizing GAI for landscape renderings.
	For instance, creativity enhancement is evident during \textit{brainstorming}, where \tool allows designers to input basic environmental requirements and incrementally introduce plant combinations without a specific scene design.
	In addition, the ability to efficiently generating landscape renderings enables users to string these renderings together in an abstract streamline and form a narrative map, as shown in Fig.~\ref{fig:narrativemap}.}
	In this novel AI-assisted design approach, the rendering, previously only present in the final result presentation, now serves as a substitute for the scene sketch and is incorporated into the initial concept development.
	We were pleased to discover that systems featuring intuitive and well-tailored interactive capabilities exhibited greater user-friendliness and were more readily embraced by designers, who could freely explore novel application scenarios.
	This realization lead us to consider that redefining the inputs and outputs of the traditional design process could serve as a starting point for envisioning a new design paradigm \rev{for creative design}, particularly in light of the efficient generative capabilities offered by AI.

\subsubsection{\rev{Method generalizability}}
\rev{
While this study centers on landscape design, the proposed pipeline and fine-tuning strategies are readily adaptable to diverse creative designs.
Breaking down an end-to-end process into a multi-step pipeline and integrating it with LLMs is applicable to other GAI-assisted designs with “iterating prompts and outputs”~\cite{vimpari2023adapt}.
Specifically, a multi-step pipeline can enhance element relation arrangement and layout adjustment in the generation process, a characteristic often seen in architectural, fashion, and UI design.
In these fields, textual descriptions alone may inadequately convey the intricacies of design intentions~\cite{ko2023large}.
Our method, incorporating graph and layout as a structured forward process and interactive approach, enables a more nuanced interpretation of design intentions, simplifying the challenge of constructing a large-scale fine-tuning dataset.
This encapsulation of domain knowledge into the pipeline proves essential for tasks where preparing a fine-tuning dataset is challenging.
}

\subsection{Limitations}\label{subsec:limitation}

\underline{Generation performance.}
While in most cases, our system is capable of generating the correct type of plant in the intended location. 
However, there are instances where certain types of plants can influence each other. 
For example, the generated plant may be of the same type even if the user input specifies a different type. 
This issue may stem from an inherent limitation of GLIGEN. In our experiment, we observed that when certain objects are generated together, one object is converted into another within the GLIGEN base model. 
To address this situation, we attempt to mitigate the effect by using latent composition and freezing the generated single object latent. 
While this approach helps maintain the correct type of generated plant, it does result in a decrease in the coherence between the object and the background.
 In future work, we aim to investigate how and when different concepts cause this conversion to avoid such conditions and improve our system's usability.

 \rev{\underline{Bias caused by model fine-tuning.}
Pre-trained models, while equipped with a broad knowledge base, can inadvertently carry biases if their training datasets lack diversity or exhibit certain preferences.
We observed that while some popular styles like Japanese Zen Gardens and English Cottage Gardens are rendered with high fidelity, others are limited to a narrow selection of plant types or even incorrect species.
Fine-tuning with LoRA is designed to customize the model's output more precisely for landscape design tasks.
However, if the fine-tuning dataset is not sufficiently diverse, the model may become overly tailored to the particular styles, plants, and materials within that dataset, at the expense of other viable and innovative design options.
This situation presents a critical trade-off: the specificity and accuracy of generated designs versus the diversity and versatility required by designers.
To balance this trade-off, the weight of LoRA model, fine-tuning dataset collection and other methods like regularization should be carefully considered in accordance with the designers' objectives.
 }

Another limitation is referred to as the \underline{concept pollution} of LoRA. 
When a LoRA model is merged with the base model, certain concepts in the base model can be altered or contaminated by the LoRA model. 
For instance, if we load a LoRA model focusing on "banya" trees into the base model, all the trees in the base model may become "banya" trees. 
To alleviate this issue, we employ regularization techniques. 
However, it is worth noting that when encountering strong features, such as colorful flowers, it is possible for the flowers in the base model to be contaminated and appear as the same type of colorful flower. 
In future work, we aim to augment the regularization dataset to prevent such concept pollution and further enhance the system's performance.

\underline{System design}.
The current system's selection of plant species is limited due to the concept pollution problem. 
Resolving this issue could enable the system to support a wider range of species.
Furthermore, plant selection in our system relies on the designer's experience for configuration, with the ecological harmony of configured plants judged solely by the designer, which raises the system's threshold. 
In the future, incorporating a graph recommendation model with expert knowledge input could enable automatic plant species recommendations and new graph generation on the graph panel, better aligning with landscape design practice.

\underline{AI-assisted system with multi-interactions}.
During the post-study interviews with designers, we learned that they are accustomed to using hand-drawn sketches and expressed their desire for a function allowing requirement transformation in the form of drawings directly on the drawing surface.
This insight highlighted the importance of considering users' habits when selecting different interaction methods with the consideration of diverse user input (text, voice, drawing, graphical control, etc.).
It can maximize the smoothness of user command transformation within the system.

\section{Conclusion and Future Work}\label{sec:conclusion}

This paper has explored the potential of integrating large pre-trained generative models within various design phases to facilitate a human-AI iterative design process.
Addressing the limitations of existing end-to-end rendering generation methods, \tool was developed through a formative study and support scene graph and layout as visual interaction components to ensure more compatibility with common design processes. 
The system incorporates a two-stage pipeline, consisting of a concretization module for translating conceptual ideas into concrete scene layouts, and an illustration module for transforming scene layouts into realistic landscape renderings. 
Performance evaluations have attested to \tool's effectiveness in landscape rendering generation. \rev{A within-subjects study comparing conventional design process with the AI-assisted design process has demonstrated the effectiveness of \tool and also uncovered potential areas for improvements.}

Future work will focus on enhancing the generation controllability of AI-assisted landscape designing tools by addressing current limitations, such as concept pollution and unexpected object type conversion. 
To augment user experience and efficiency, our system will also be equipped with supplementary features, such as plant pairing.
Building upon this research foundation, we aim to delve deeper into novel design paradigms that foster collaboration between artificial intelligence and human designers. 
Additionally, we intend to investigate effective interaction patterns and pertinent information circuits tailored for the enhancement of future AI-supported design processes.


\appendix
\newpage
\section{Glossary}


\printnoidxglossary[style=twocolborderbold, title={\fontsize{16}{22}}]

\end{document}